\documentclass[amsmath,amssymb,aps, twocolumn, pre, nofootinbib]{revtex4-1}
\pdfoutput=1
\usepackage[utf8]{inputenc}
\usepackage[T1]{fontenc}
\usepackage{multirow}
\usepackage{graphicx}
\graphicspath{ {Images/} }
\usepackage{booktabs}
\usepackage{caption}
\usepackage{natbib}
\usepackage{float}
\usepackage{textcase}
\usepackage{svg}
\usepackage{url} 
\usepackage{bbm}

\usepackage{hhline}
\usepackage{array}
\newcolumntype{P}[1]{>{\centering\arraybackslash}p{#1}}
\begin{document}
	
	\title{Identification of skill in an online game: The case of Fantasy Premier League}
	\author{Joseph D.~O'Brien}
	\author{James P.~Gleeson}
	\author{David J.~P.~O'Sullivan}
	\affiliation{MACSI, Department of Mathematics and Statistics, University of Limerick, Limerick V94 T9PX, Ireland}
	\date{\today}
	
	
	
	\begin{abstract}
		In all competitions where results are based upon an individual's performance the question of whether the outcome is a consequence of skill or luck arises. 
		We explore this question through an analysis of a large dataset of approximately one million contestants playing \textit{Fantasy Premier League}, an online fantasy sport where managers choose players from the English football (soccer) league. 
		We show that managers' ranks over multiple seasons are correlated and we analyse the actions taken by managers to 
		increase their likelihood of success. The prime factors in determining a manager's success are found to be long-term planning and consistently good decision-making in the face of the noisy contests upon which this game is based. Similarities between managers' decisions over time that result in the emergence of `template' teams, suggesting a form of herding dynamics taking place within the game, are also observed. Taken together, these findings indicate common strategic considerations and consensus among successful managers on crucial decision points over an extended temporal period.
	\end{abstract}

	\maketitle
	\section{Introduction}
	
	Hundreds of millions of people consume sporting content each week, motivated by several factors. These motivations include the fact that the spectator enjoys both the quality of sport on display and the feeling of eustress arising from the possibility of an upset \cite{mumford2013watching, wann1995preliminary}. This suggests that there are two important elements present in sporting competition: a high level of skill among players that provides aesthetic satisfaction for the spectator and also an inherent randomness within the contests due to factors such as weather, injuries, and in particular luck. The desire for consumers to get further value from their spectating of sporting content has resulted in the emergence of \emph{fantasy sports} \cite{lee2013understanding, dwyer2011love, karg2011fantasy, farquhar2007types}, in which the consumers, or \emph{managers} as we shall refer to them throughout this article, begin the season with a virtual budget from which to build a team of \emph{players} who, as a result of partaking in the real physical games, receive points based upon their statistical performances. The relationship between the fantasy game and its physical counterpart raises the question of whether those who take part in the former suffer (or gain) from the same combination of skill and luck that makes their physical counterpart enjoyable.
	
	
	The emergence of large scale quantities of detailed data describing the dynamics of sporting games has opened up new opportunities for quantitative analysis, both from a team perspective \cite{park2005network,yamamoto2011common,grund2012network,gabel2012random, ribeiro2016advantage, gudmundsson2017spatio, gonccalves2017exploring, buldu2019defining} and also at an individual level~\cite{onody2004complex,saavedra2010mutually, duch2010quantifying, radicchi2011best, mukherjee2012identifying, cintia2013engine, brooks2016developing}. This has resulted in analyses aiming to determine two elements within the individual sports; firstly quantifying the level of skill in comparison to luck in these games \cite{yucesoy2016untangling, ben2013randomness, grund2012network, aoki2017luck, pappalardo2018quantifying} while, secondly, identifying characteristics that suggest a difference in skill levels among the competing athletes \cite{duch2010quantifying,radicchi2012universality}. Such detailed quantitative analysis is not, however, present in the realm of fantasy sports, despite their burgeoning popularity with an estimated 45.9 million players in the United States alone in 2019 \cite{fsgaDemographics}. One notable exception is a recent study~\cite{getty2018luck}, which derived an analytical quantity to determine the role chance plays in many contests including fantasy sports based on American sports, and suggested that skill was a more important factor than luck in the games.
	
	
	\begin{figure*}[t]
		\centering
		\includegraphics[width = \textwidth]{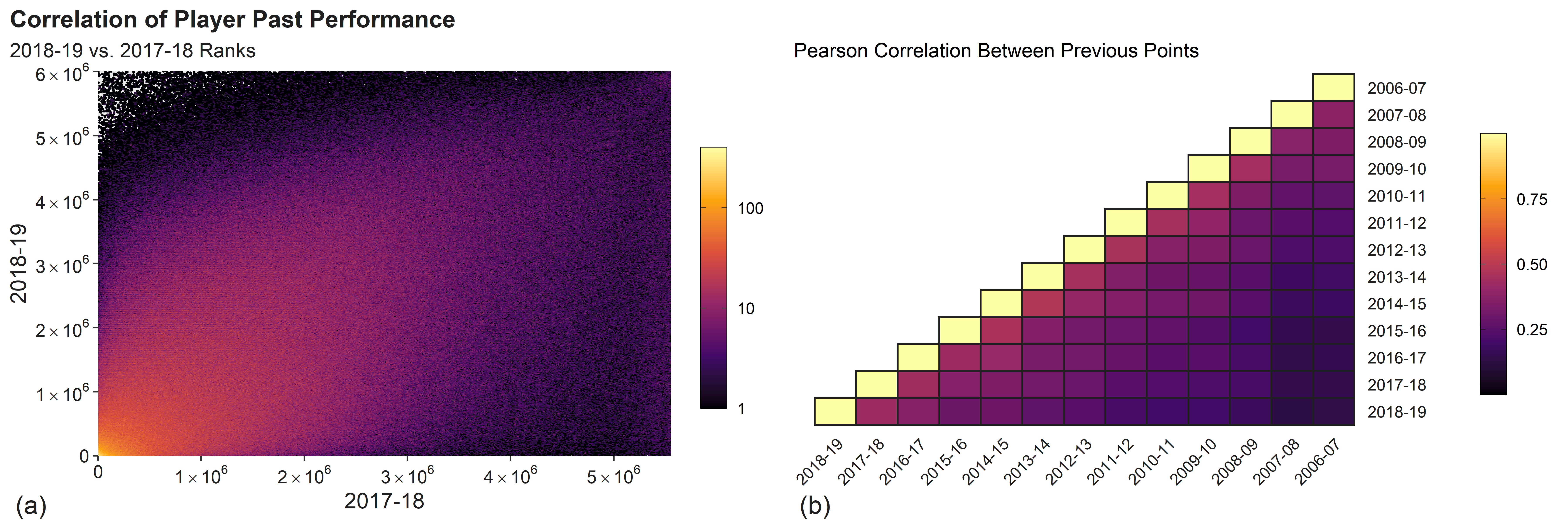}
		\caption{\textbf{Relationship between the performance of managers over
		seasons of FPL}. (a) The relationship between managers' ranks in the 2018/19 and 2017/18 seasons. Each bin is of width 5,000 with the colour highlighting the number of managers in each bin; note the logarithmic scale in colour. (b) The pairwise Pearson correlation between a manager's points totals over multiple seasons of the game, calculated over all managers who appeared in both seasons.}
		\label{fig:history_plots}
	\end{figure*}
	
	Motivated by this body of work, we consider a dataset describing the \textit{Fantasy Premier League (FPL)} \cite{fplsite}, which is the online fantasy game based upon the top division of England's football league. This game consists of over seven million \textit{managers}, each of whom builds a virtual team based upon real-life players. Before proceeding, we here introduce a brief summary of the rules underlying the game, to the level required to comprehend the following analysis~\cite{fplRules}. The (physical) Premier League consists of 20 teams, each of whom play each other twice, resulting in a season of 380 fixtures split into 38 unique \textit{gameweeks}, with each gameweek generally containing ten fixtures. A manager in FPL has a virtual budget of \pounds100m at the initiation of the season from which they must build a squad of 15 players from the approximately 600 available. Each player's price is set initially by the game's developers based upon their perceived value to the manager within the game, rather than their real-life transfer value. The squad of 15 players is composed under a highly constrained set of restrictions which are detailed in~\ref{app:rules}. 
	
	In each gameweek the manager must choose 11 players from their squad as their team for that week and is awarded a points total from the sum of the performances of these players (see~\ref{tab:points}). The manager also designates a single player of the 11 to be the captain, with the manager receiving double this players' points total in that week. Between consecutive gameweeks the manager may also make one unpenalised change to their team, with additional changes coming as a deduction in their points total. The price of a given player then fluctuates as a result of the supply-and-demand dynamic arising from the transfers across all managers' rosters. The intricate rules present multiple decisions to the manager and also encourages longer-term strategising that factors in team value, player potential, and many other elements.

	In Section \ref{subsec:hist_corr} we analyse the historical performance of managers in terms of where they have ranked within the competition alongside their points totals in multiple seasons, in some cases over a time interval of up to thirteen years. We find a consistent level of correlation between managers' performances over seasons, suggesting a persistent level of skill over an extended temporal scale. 
	Taking this as our starting point, in Section \ref{subsec:specific_season} we aim to understand the decisions taken by managers which are indicative of this skill level over the shorter temporal period of the 38 gameweeks making up the 2018/19 season by analysing the entire dataset of actions taken by the majority of the top one million managers\footnote{Due to data availability issues at the time of collection such as managers not taking part in the entire season, the final number of managers identified was actually 901,912. We will however, for the sake of brevity, refer to these as the top 1 million managers over the course of this article. It is also important to note that data from previous seasons is unattainable, which is why we restrict this detailed study to the 2018/19 season.} over the course of the season. Even at this shorter scale we find consistent tiers of managers who, on a persistent basis, outperform those at a lower tier. 
	
	With the aim of identifying why these differences occur, we present (Section \ref{subsec:decisions}) evidence of consistently good decision making with regard to team selection and strategy. This would be consistent with some common form of information providing these skilled managers with an `edge' of sorts, for example in the US it has been suggested that 30\% of fantasy sports participants take advantage of further websites when building their teams~\cite{burke2016exploring}. Arguably most interesting of all, in Section~\ref{subsec:Template} we demonstrate how at points throughout the season there occurs temporary herding behaviour in the sense that managers appear to converge to consensus on a \textit{template team}. However, 
	the consensus does not persist in time, with managers subsequently differentiating themselves from the others. We consider possible reasons and mechanisms for the emergence of these template teams. 
	
	\section{Results}
	
	\begin{figure*}[t]
		\centering
		\includegraphics[width = \textwidth]{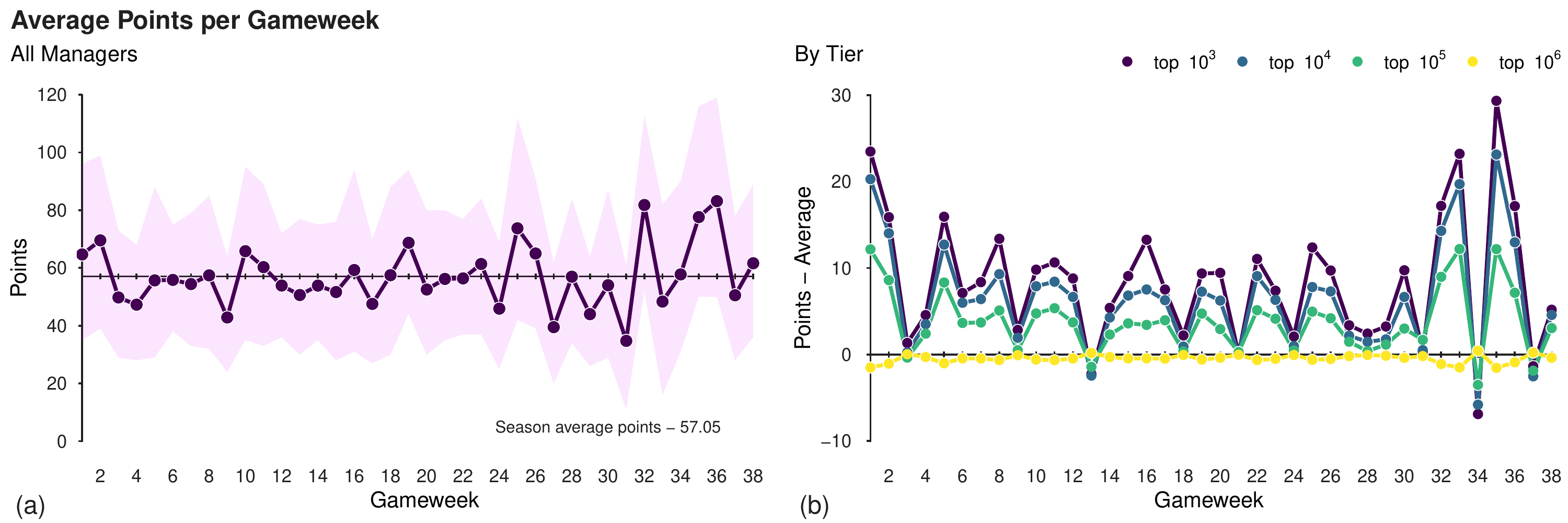}
		\caption{\textbf{Summary of points obtained by managers over the course
				of the 2018/19 season.} (a) The mean number of points over all managers for each GW. The shaded regions denote the 95\% percentiles of the points' distribution. (b) The difference between the average number of points for four disjoint tiers 
			of manager, the top $10^3, 10^4, 10^5,\text{ and } 10^6$, and the overall average points as per panel (a). Note that managers are considered to be in only one tier so, for example, the top-$10^4$ tier contains managers ranked from 1001 to $10^4$.}
		\label{fig:ave_points_class}
	\end{figure*}
	
	\subsection{Historical Performance of Players}\label{subsec:hist_corr}
	
	We consider two measures of a manager's performance in a given season of FPL: the total number of points their team has obtained over the season and also their resulting rank based on this points total in comparison to all other managers. A strong relationship between the managers' performances over multiple seasons of the game is observed. For example, in panel (a) of Fig.~\ref{fig:history_plots} we compare the ranks of managers who competed in both the 2018-19 and 2017-18 seasons. The density near the diagonal of this plot suggests a correlation between performances in consecutive seasons. Furthermore, we highlight specifically the bottom left corner which indicates that those managers who are among the most highly ranked appear to perform well in both seasons. Importantly, if we consider the top left corner of this plot it can be readily seen that the highest performing managers in the 2017-18 season, in a considerable number of cases, did not finish within the lowest positions in the following season as demonstrated by the speckled bins with no observations.
	
	This is further corroborated in panel (b), in which we show the pairwise Pearson correlation between the total points obtained by managers from seasons over a period of 12 years. While the number of managers who partook in two seasons tends to decrease with time, a considerable number are present in each comparison. Between the two seasons shown in Fig.~\ref{fig:history_plots}(a) for example, we observe results for approximately three million managers and find a correlation of 0.42 among their points totals. Full results from 13 consecutive seasons, including the number managers present in each pair and the corresponding Pearson correlation coefficients, are given in \ref{table:correlations}.
	
	Using a linear regression fit to the total points scored in the 2018/19 season as a function of the number of previous seasons in which the manager has played (\ref{fig:points_history}) we find that each additional year of experience is worth on average 22.1 ($\text{R}^2 = 0.082$) additional points (the overall winner in this season obtained 2659 points). This analysis suggests that while there are fluctuations present in a manager's performance during each season of the game, there is also some consistency in terms of performance levels, suggesting a combination of luck and skill being present in fantasy sports just as was observed in their physical analogue in \cite{getty2018luck}.
	
	\subsection{Focus on Season 2018-19}\label{subsec:specific_season}
	
	In Sec.~\ref{subsec:hist_corr} we considered, over multiple seasons, the performance of managers at a season level in terms of their cumulative performance over the 38 gameweeks of each season. We now focus at a finer time resolution, to consider the actions of managers at the gameweek level for the single season 2018/19, in order to identify elements of their decision making which determined their overall performance in the game. 
	
	The average points earned by all managers throughout the season is shown in Fig.~\ref{fig:ave_points_class}(a) along with the 95 inter-percentile range, i.e., the values between which the managers ranked in quantiles 0.025 to 0.975 appear.
	This quantity exhibits more frequent fluctuations about its long-term average (57.05 points per gameweek) in the later stages of the season, suggesting that some elements of this stage of the season cause different behaviour in these gameweeks. There may of course be many reasons for this e.g., difficult fixtures or injuries for generally high-scoring players or even simply a low/high scoring gameweek, which are themselves factors of luck within the sport itself (see \ref{tab:manager_points} for a detailed break down of points per gameweek). However, in Section \ref{subsec:chips} we analyse an important driver of the fluctuations related to strategic decisions of managers in these gameweeks.
	
	In each season some fixtures must be rescheduled due to a number of reasons, e.g., clashing fixtures in European competitions, which results in certain gameweeks that lack some of the complete set of ten fixtures. Such scenarios are known as \textit{blank-gameweeks} (BGW) and their fixtures are rescheduled to another gameweek in which some teams play twice; 
	these are known as \textit{double-gameweeks} (DGWs). In the case of the 2018/19 season these BGWs took place in GWs 27 (where there were eight fixtures), 31 (five fixtures), and 33 (six fixtures), making it difficult for some managers to have 11 starting players in their team. The DGWs feature some clubs with two games and therefore players in a manager's team who feature in these weeks will have twice the opportunity for points; in the 2018/19 season these took place in GWs 25 (where 11 games were played), 32 (15), 34 (11), and 35 (14). We see that the main swings in the average number of points are actually occurring in these gameweeks (aside from the last peak in GW 36 which we will comment on later in the article). In Section \ref{subsec:chips} we show that the managers' attitude and preparation towards these gameweeks are in fact indicators of their skill and ability as a fantasy manager.
	
	\begin{figure}
		\centering
		\includegraphics[width = 0.48\textwidth]{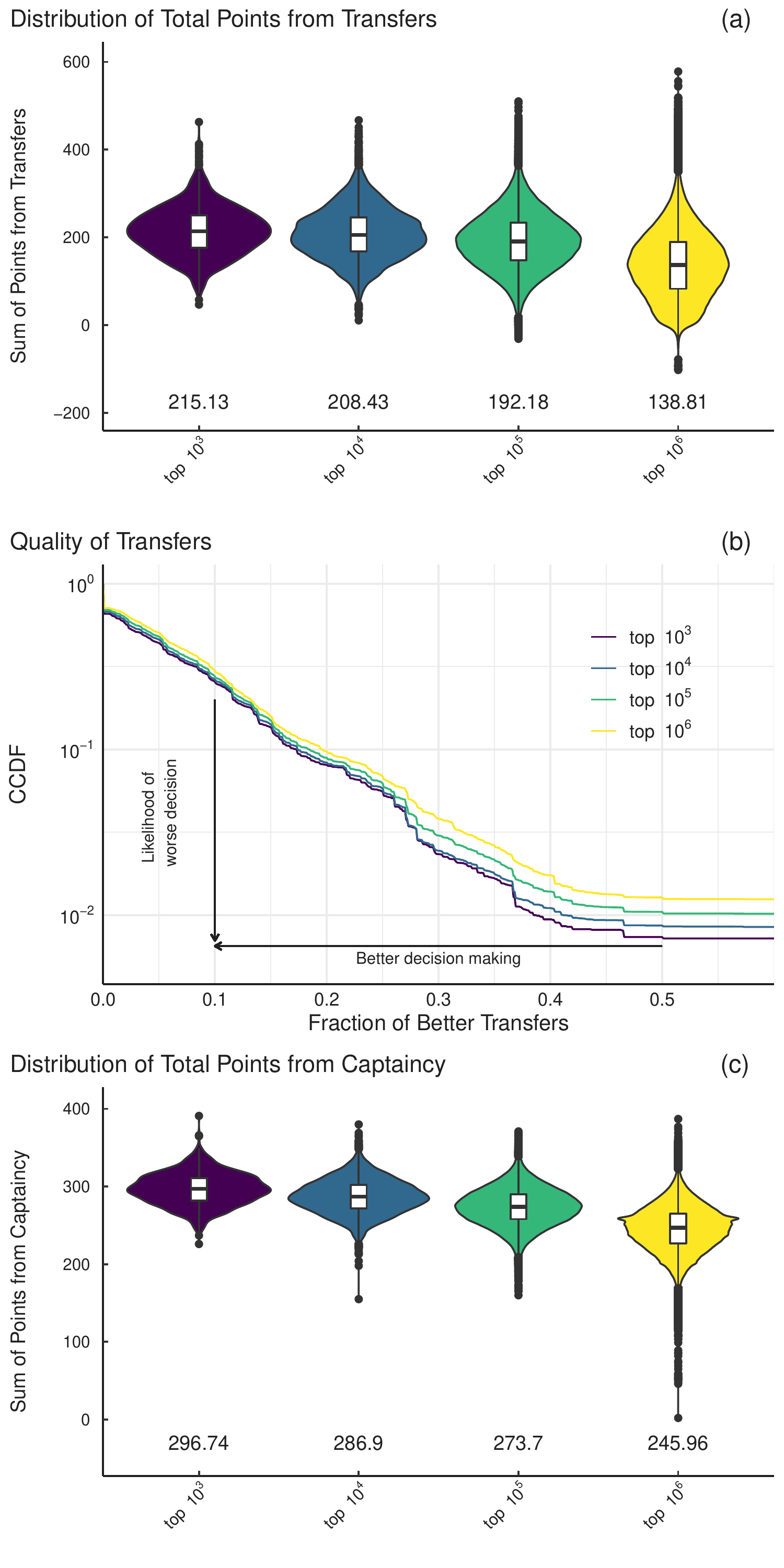}
		\caption{\textbf{Decisions of managers by tier.}
			(a)~Distributions of the total net points earned by managers in the gameweek following a transfer, i.e., the points scored by the player brought in minus that of the player transferred out. The average net points for each tier is also shown below; note the difference between the top three tiers and the bottom tier. (b)~Distribution of the fraction of better transfers a manager could have made based upon points scored in the following gameweek. Faster-decreasing distributions reflect managers in that tier being more successful with their transfers. (c) The distribution of points from captaincy along with the average total for each tier.} 
		\label{fig:decision_plots}
	\end{figure}
	
	To analyse the impact of decision-making upon final ranks, we define \emph{tiers} of managers by rank-ordering them by their final scores and then splitting into the top $10^3$, top $10^4$, top $10^5$, and top $10^6$ positions. These disjoint
	tiers of managers, i.e., the top $10^3$ is the managers with ranks between 1 and 1000, the top $10^4$ those with ranks between 1001 and 10,000 and so on, range from the most successful (top $10^3$) to the relatively unsuccessful (top~$10^6$) and so provide a basis for comparison (see \ref{tab:manager_points} and \ref{table:points_summary} for summaries of points obtained by each tier).
	The average performance of the managers in each tier (relative to the baseline average over the entire dataset) are shown in panel (b) of Fig.~\ref{fig:history_plots}. Note that the points for the top~$10^6$ tier are generally close to zero as the calculation of the baseline value is heavily dependent upon this large bulk of managers. A detailed summary of each tier's points total, along with visualisation of the distribution of points total may be found in \ref{tab:points} and \ref{fig:points_summary}. It appears that the top tier managers outperform those in other tiers, not only in specific weeks but consistently throughout the season which results in the competition for places in this top tier more difficult to obtain as the season progresses (\ref{fig:alluvial_plots}). This is particularly noticeable in the first gameweek, where the top $10^3$ managers tended to perform very strongly, suggesting a high level of preparation (in terms of squad-building) 
	prior to the physical league starting. We also comment that the largest gaps between the best tier and the worst tier occur not only in two of the special gameweeks (DGW 35 and BGW 33) but also in GW 1, which suggests that prior to the start of the season these managers have built a better-prepared team to take advantage of the underlying fixtures. We note however that all tiers show remarkably similar temporal variations in their points totals, in the sense that they all experience simultaneous peaks and troughs during the season. See \ref{tab:manager_points} for a full breakdown of these values alongside their variation for each gameweek.
	
	Having identified both differences and similarities underlying the performance in terms of total points for different tiers of managers we now turn to analysis of the actions that have resulted in these dynamics.
	
	\subsection{Decision-Making}\label{subsec:decisions}
	\subsubsection{Transfers}
	The performance of a manager over the season may be viewed as the consequence of a sequence of decisions that the manager made at multiple points in time. These decisions include which players in their squad should feature in the starting team, the formation in which they should set up their team, and many more. In the following sections we consider multiple scenarios faced by managers and show that those who finished within a higher tier tended to consistently outperform those in lower tiers.
	
	One decision the manager must make each gameweek is whether to change a player in their team by using a transfer. If the manager wants to make more than one transfer they may also do so but at the cost of a points deduction for each extra transfer. The distribution of total points made from transfers, which we determine by the difference between points attained by the player the manager brought in for the following gameweek compared to the player whom they transferred out, over the entire season for each tier is shown in Fig.~\ref{fig:decision_plots}(a). The average number for each tier is also shown. To further analyse this scenario we calculate, for each gameweek, the number of better transfers the managers could have made with the benefit of perfect foresight, given the player they transferred out. This involves taking all players with a price less than or equal to that of the player transferred out and calculating the fraction of options which were better than the one selected, i.e., those who received more points the following gameweek (see Methods). Figure~\ref{fig:decision_plots}(b) shows the complementary cumulative distribution function (CCDF) of this quantity for each tier, note the steeper decrease of the CCDFs for the higher tiers implies that these managers were more likely to choose a strong candidate when replacing a player.
	
	\begin{figure}
		\centering
		\includegraphics[width = 0.48\textwidth]{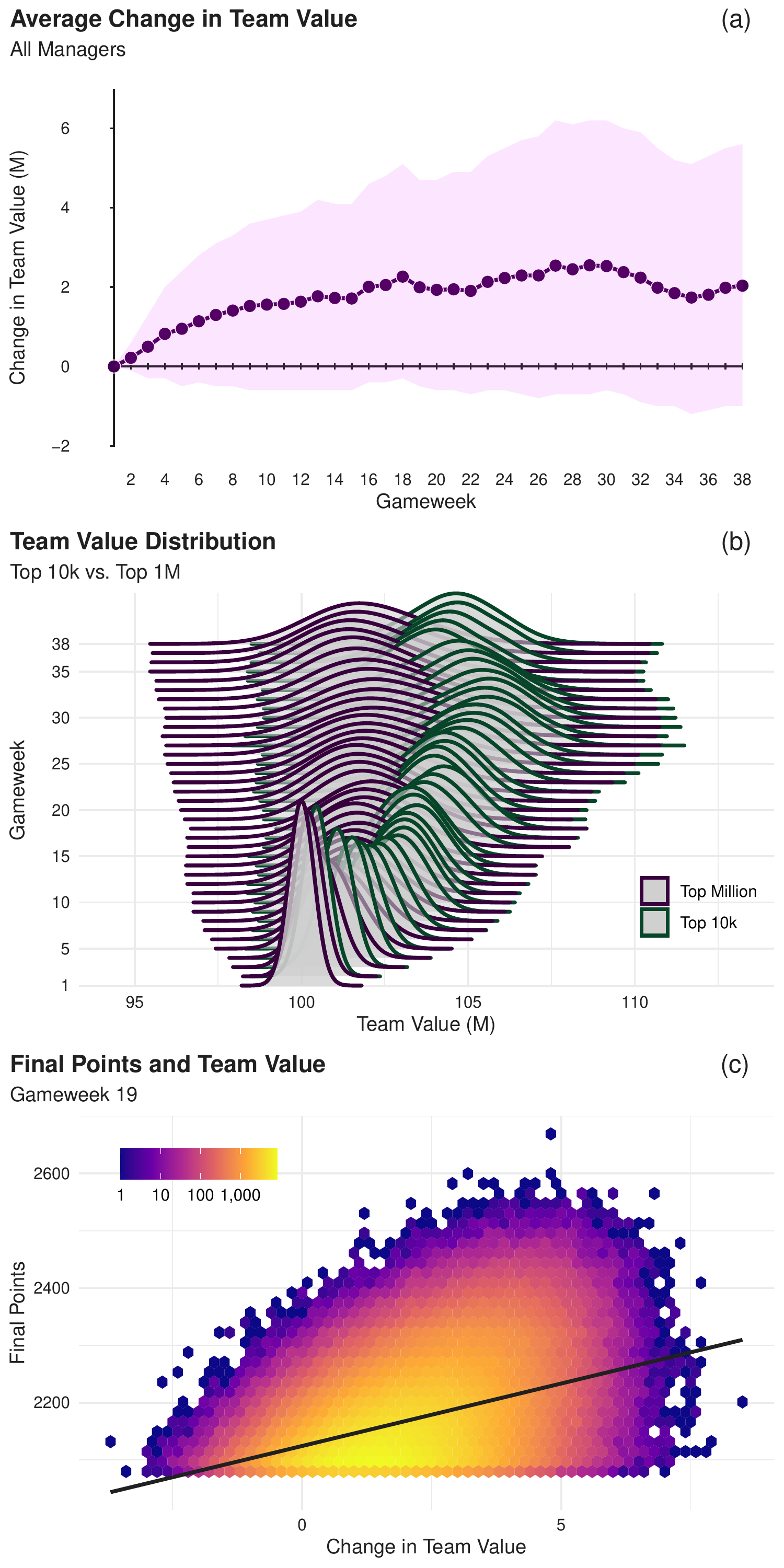}
		\caption{\textbf{Analysis of the team value of managers.} (a) The change in average team value from the initial \pounds100M of all managers, along with 95 percentiles; note the general upward trend of team value over the course of the season. (b)~Distributions of team values for each gameweek for those who finished in the top ten thousand positions (i.e., the combination of those in the top $10^3$ and $10^4$ tiers) versus lower-ranked managers. The distribution for those with higher rank is generally to the right of that describing the other managers from an early stage of the season, indicating higher team value being a priority for successful managers. (c) The relationship between a manager's team value at GW 19 versus their final points total, where the heat map indicates the number of managers within a given bin. The black line indicates the fitted linear regression line, showing that an increase in team value by \pounds1M at this point in the season results in an average final points increase of 21.8 points.} 
		\label{fig:money_plots}
	\end{figure}
	
	\begin{figure*}[t]
		\centering
		\includegraphics[width = \textwidth]{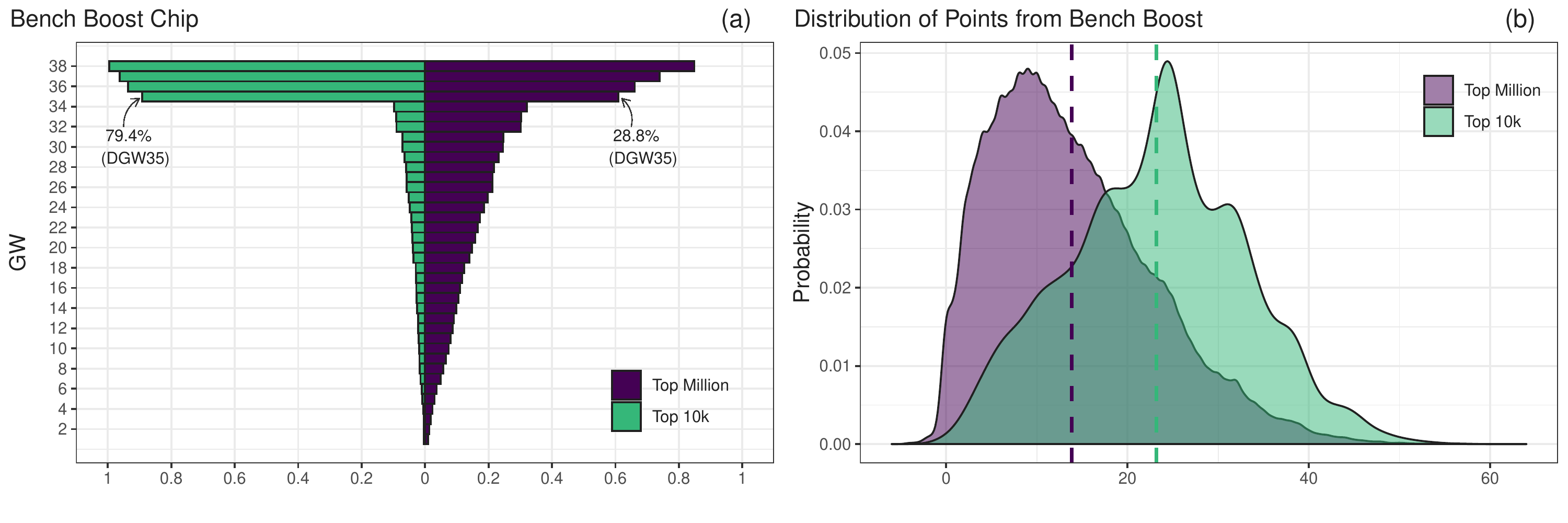}
		\caption{\textbf{Summary of use and point returns of the bench boost chip.} The managers are grouped into two groups: those who finished in the top ten-thousand positions (Top 10k) and the remainder (Top Million). (a) Fraction of managers who had used the bench boost chip by each gameweek. We see a clear strategy for use in double gameweek 35, particularly for the top managers, 79.4\% of whom used it at this stage. (b) Distribution of points earned from using this chip along with the average points---23.2 for the Top 10k and 13.8 for the Top Million---shown by the dashed lines.} 
		\label{fig:bb_usage}
	\end{figure*}
	
	A second decision faced by managers in each gameweek is the choice of player to nominate as captain, 
	which results in the manager receiving double points for this players' actions during the GW. This is, of course, a difficult question to answer as the points received by a player can be a function of both their own actions i.e., scoring or assisting a goal, and also their team's collective performance (such as a defender's team not conceding a goal).
	This topic is an identification question which may be well suited to further research making use of the data describing the players and teams but with additional data about active managers who are making the same decision. For example, an analysis of the captaincy choice of managers based upon their social media activity was recently presented in \cite{bhatt2019should} and showed that the \textit{wisdom of crowds} concept performs comparably to that of the game's top managers. Panel (c) of Fig.~\ref{fig:decision_plots} shows the distribution of points obtained by managers in each tier from their captaincy picks. Again we observe that the distribution of points obtained over the season is generally larger for those managers in higher tiers.

	\subsubsection{Financial Cognizance}

	The financial ecosystem underlying online games 
	has been a focus of recent research 
	\cite{papagiannidis2008making, yamaguchi2004analysis}. With this in mind, we consider the importance of managers' financial awareness in impacting their performance. As mentioned previously, each manager is initially given a budget of \pounds100 million to build their team, constrained by the prices of the players which, themselves fluctuate over time. While the dynamics of player price changes occur via an undisclosed mechanism, attempts to understand this process within the community of Fantasy Premier League managers have resulted in numerous tools to help managers predict player price changes during the season, for example see \cite{fplstats}. The resulting algorithms are in general agreement that the driving force behind the changes is the supply and demand levels for players.

	These price fluctuations offer an opportunity for the astute manager to `play the market' and achieve a possible edge over their rivals and allow their budget to be more efficiently spent (see \ref{fig:player_points} for a description of player value and their corresponding points totals and \ref{fig:ternary} for an indication of how the managers distribute their budget by player position). At a macro level this phenomenon of price changes is governed by the aforementioned supply and demand, but these forces are themselves governed by a number of factors affecting the player including, but not limited to, injuries, form, and future fixture difficulty. As such, managers who are well-informed on such aspects may profit from trading via what is in essence a fundamental analysis of players' values by having them in their team prior to the price rises \cite{dechow2001short}. Interestingly, we note that the general trend of team value is increasing over time among our managers as shown in panel (a) of Fig~\ref{fig:money_plots} along with corresponding 95 percentiles of the distribution, although there is an indicative decrease between weeks towards the season's end (GWs 31-35) suggesting the team value becomes less important to the managers towards the games conclusion. Equivalent plots for each tier are shown in \ref{fig:tv_class}.
	
	Probing further into the relationship between finance and the managers' rank, we show in Fig.~\ref{fig:money_plots}(b) the distribution of team values for the top two tiers (top $10^3$ and top $10^4$), compared with that for the bottom two tiers (top $10^5$ and top $10^6$) 
	There is a clear divergence between the two groups from an early point in the season, indicating an immediate importance being placed upon the value of their team. A manager who has a rising team value is at an advantage relative to one who does not due to their increased purchasing power in the future transfer market. This can be seen in panel (c) of Fig.~\ref{fig:money_plots} which shows the change in team value for managers at gameweek 19, the halfway point of the season, versus their final points total. A positive relationship appears to exist and this is validated by fitting an OLS Linear Regression with a slope of 21.8 ($R^2 =~0.1689$), i.e., an increase of team value by \pounds1M at the halfway point is worth, on average, an additional 21.8 points by the end of the game (for the same analysis in other gameweeks see \ref{tab:regression_coefficients}). The rather small $R^2$ value suggests, however, that the variation in a managers' final performance is not entirely explained by their team value and as such we proceed to analyse further factors which can play a part in their final ranking.

	
	\subsubsection{Chip Usage}\label{subsec:chips}

	A further nuance to the rules of FPL is the presence of four \textit{game-chips}, which are single use `tricks' that may be used by a manager in any GW to increase their team's performance, by providing additional opportunities to obtain points. The time at which these chips are played and the corresponding points obtained are one observable element of a managers' strategy. 
	A detailed description for each of the chips and analysis of the approach taken by the managers in using them is given in \ref{sm:chips}.
	
	For the sake of brevity we focus here only on one specific chip, the \textit{bench boost}. When this chip is played, the manager receives points from all fifteen players in their squad in that GW, rather than only the starting eleven as is customary. This clearly offers the potential for a large upswing in points if this chip is played in an efficient manner, and as such it should ideally be used in GWs where the manager may otherwise struggle to earn points with their current team or weeks in which many of their players have a good opportunity of returning large point scores. The double and blank GWs might naively appear to be optimal times to deploy this chip
	however when the managers' actions are analysed we see differing approaches (and corresponding returns).
	
	Figure~\ref{fig:bb_usage} shows the proportion of managers who had used the bench boost chip by each GW alongside the corresponding distribution of points the manager received from this choice, where we have grouped the two higher tiers into one group and the remaining managers in another for visualization purposes (see \ref{fig:chip_usage} \& \ref{fig:wc_use} and \ref{tab:BB}-\ref{tab:WC} for a breakdown of use and point returns by each tier). It is clear that the majority of better performing managers generally focused on using these chips during the double and blank GWs with 79.4\% choosing to play their BB chip during DGW35 in comparison to only 28.9\% of those in the rest of the dataset. We also observe the difference in point returns as a result of playing the chip, with the distribution for the top managers being centred around considerable higher values, demonstrating that their squads were better prepared to take advantage of this chip. The fact that the managers were willing to wait until one of the final gameweeks is also indicative of the long-term planning that separates them from those lower ranked. Similar results can be observed for the other game-chips (\ref{tab:FH}-\ref{tab:WC}). We also highlight that a large proportion of managers made use of other chips in GW36, which was the later gameweek in which there was a large fluctuation from the average shown in Fig.~\ref{fig:ave_points_class}.
	
	\begin{figure*}[]
		\centering
		\includegraphics[width=0.9\textwidth]{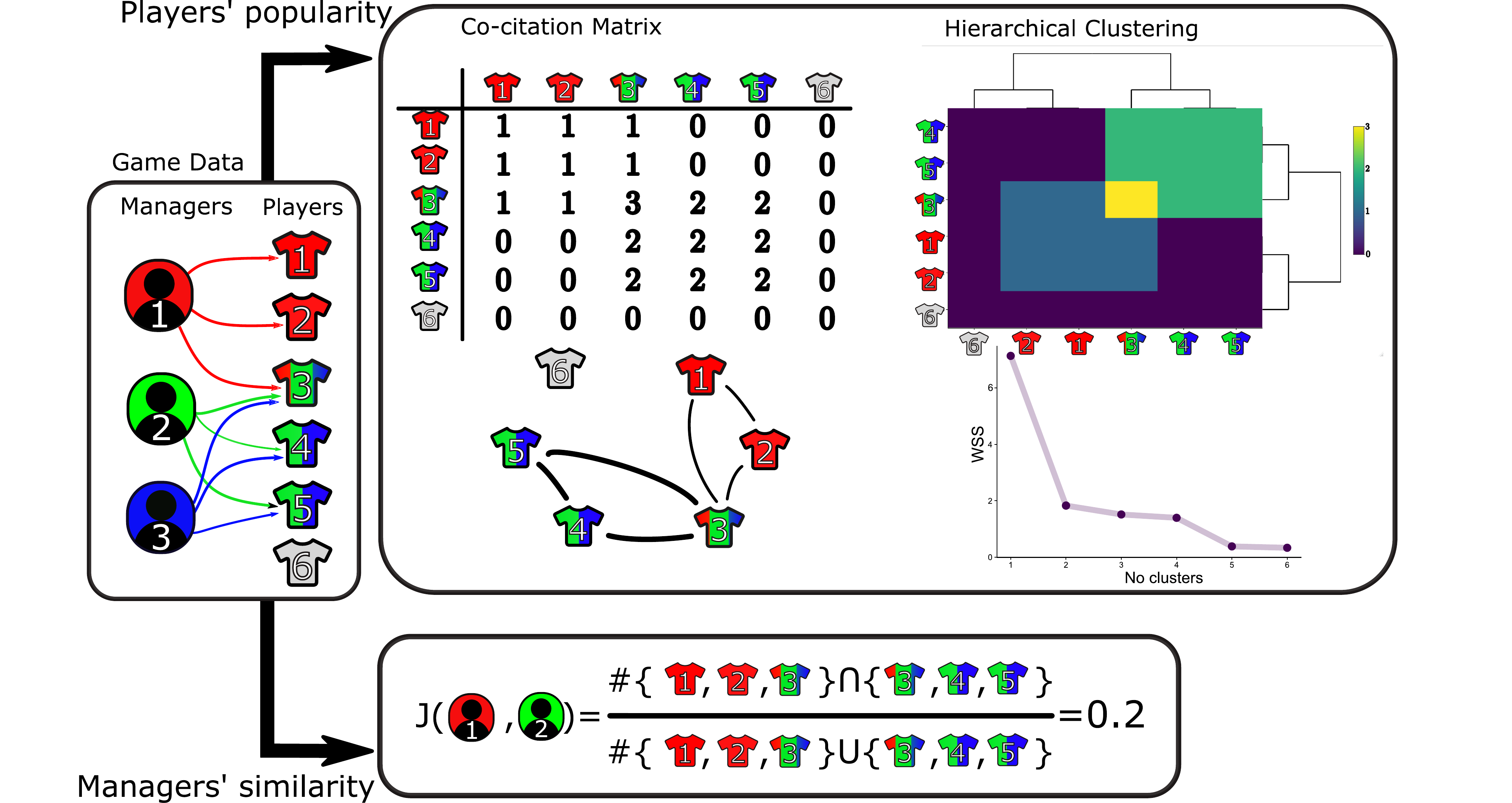}
		\caption{\textbf{Schematic representation of the approaches taken to identify similarity between the composition of managers' teams in each GW. }
			We view the connections between managers and players as a bipartite network such that an edge exists if the player is in the managers' team. To determine the relationship between players' levels of popularity we use the co-occurrence matrix which has entries corresponding to the number of teams in which two players co-appear. Using this matrix we perform hierarchical clustering techniques to identify groups of players who are similarly popular within the game, where the number of clusters is determined by analysing the within-cluster sum of squared errors. The similarity between the teams of two managers is determined by calculating the Jaccard similarity, which is determined by the number of players that appear in both teams.} 
		\label{fig:clustering_scheme}
	\end{figure*}
	
	
	Finally, we comment on the fact that some managers did not employ their chips by the game's conclusion which suggests that either they were not aware of them or, more likely, the mangers in question had simply lost interest in the game at this point. As such, the quantity of managers who had not used their chip gives us a naive estimation of the retention rate for active managers in Fantasy Premier League ($85.05\%$ of managers in our dataset). We note that this is a biased estimate in the sense that our dataset is only considering the top tiers of managers, or at least those who finished in the top tiers, and one would expect the drop out rate to be in fact much higher in lower bands. 
	
	\subsection{Template Team}\label{subsec:Template}

	While the preceding analysis proposes reasons for the differences between points obtained by tiers shown in Fig.~\ref{fig:ave_points_class}, the question remains as to why the managers'
	gameweek points totals show similar temporal dynamics.
	In order to understand this we consider here the underlying structure of the managers' teams. We show that a majority of teams feature a core group of players that results in a large proportion of teams having a similar make-up. We call this phenomenon the \textit{template team} which appears to emerge at different points in the season; this type of collective behaviour has been observed in such social settings previously, see, for example \cite{ross2014social, aleta2019dynamics}. We identify the template team by using the network structure describing the teams of all managers, which is described by the adjacency matrix $A_{ij}^G$, whereby an edge between two players $i$ and $j$ appearing in $n$ teams for a given gameweek $G$ describes a value in the matrix given by $A_{ij}^G = n$. This matrix is similar in nature to the co-citation matrix used within the field of bibliometrics \cite{newman2018networks}, see Fig.~\ref{fig:clustering_scheme} for a representation of the process.
	
\begin{figure}
		\centering
		\includegraphics[width = 0.48\textwidth]{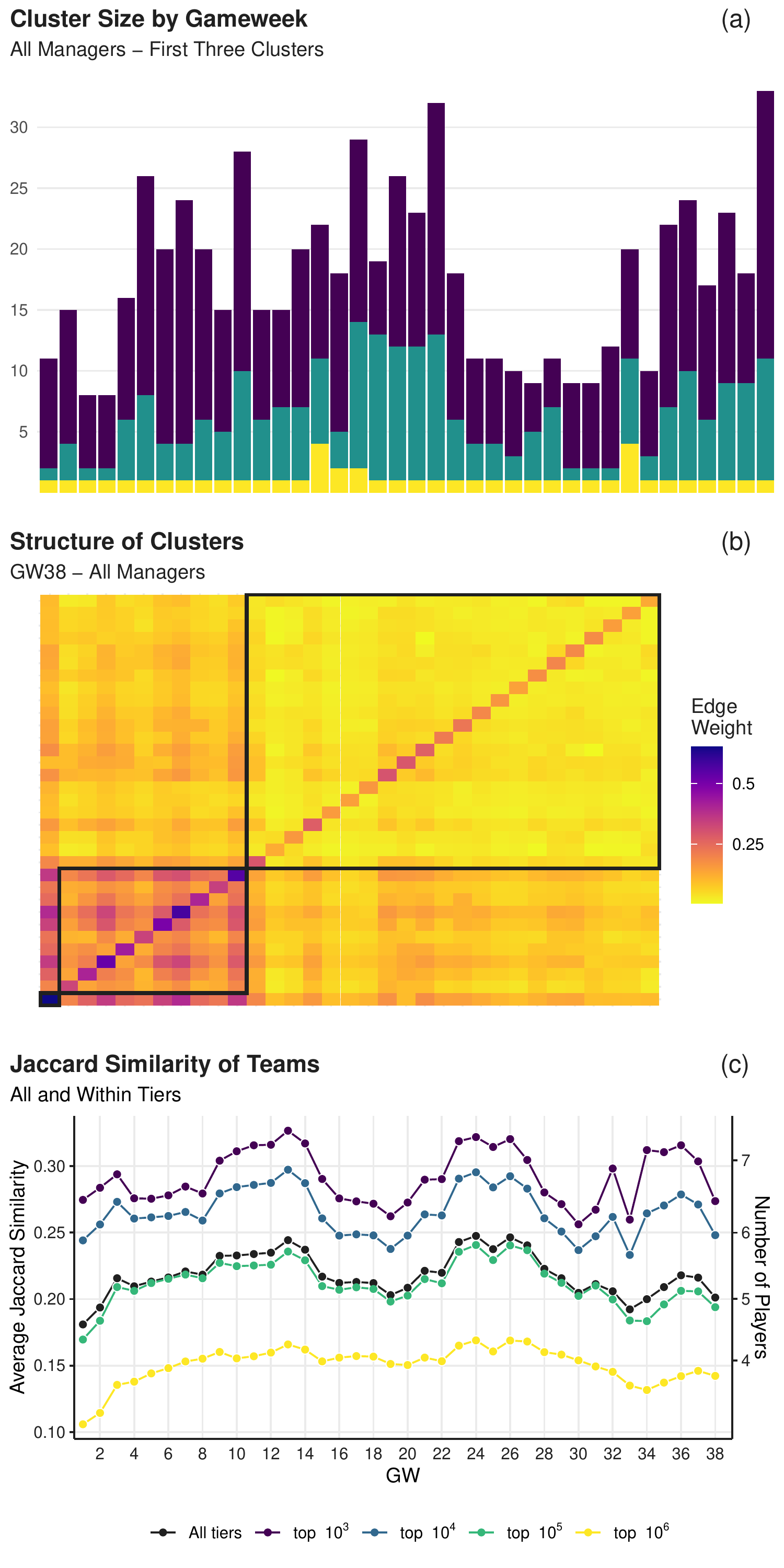}
		\caption{\textbf{Analysis of team similarity of managers.} (a) Size of each of the first three identified clusters over all managers for each gameweek. Note that the first cluster is generally of size one, simply containing the most-owned player in the game. (b)~An example of the network structure of these three clusters for gameweek 38, where we can see the ownership level decreasing in the larger clusters. The diagonal elements of this structure are the fraction of teams in which the player is present. (c)~The Jaccard similarity between the tiers of managers and also over all managers; note that the higher-performing managers tend to be more like one another than those in lower tiers, also note the fluctuations in similarity over the course of the season indicating that a template team emerges at different time points.
		} 
		\label{fig:team_similarity}
	\end{figure}

	With these structures in place we proceed to perform hierarchical clustering on the matrices in order to identify groups of players constituting the common building blocks of the managers' teams. By performing the algorithm with $k = 4$ clusters we find that three clusters contain only a small number of the 624 players, suggesting that most teams include this small group of core players (see \ref{table:first_cluster} for the identities of those in the first cluster each gameweek). Figure~\ref{fig:team_similarity}(a) shows the size of these first three clusters over all managers for each gameweek of the season (\ref{fig:clusters_all} shows the equivalent values for each tier). To understand this result further, consider that at their largest these three clusters only consist of 5.13\% (32/624) of the available players in the game, highlighting that the teams are congregated around a small group of players. For an example representation of this matrix alongside its constituent clusters we show the structure in panel (b) of Fig.~\ref{fig:team_similarity} for gameweek 38, which was the point in time at which the three clusters were largest.
	
	To further examine the closeness between managers' decisions we consider the Jaccard similarity between sets of teams, which is a distance measure that considers both the overlap and also total size of the sets for comparison (see Methods for details). Figure~\ref{fig:team_similarity}(c) shows the average of this measure over pairwise combinations of managers from all tiers and also between pairs of managers who are in the same tier.
	Fluctuations in the level of similarity over the course of the season can be seen among all tiers indicating times at which teams become closer to a template followed by periods in which managers appear to differentiate themselves more from the peers. Also note that the level of similarity between tiers increases with rank suggesting that as we start to consider higher performing managers, their teams are more like one another not only at certain parts of the season but, on average, over its entirety (see \ref{fig:jaccard_all} for corresponding plots for each tier individually). The high level of similarity between the better managers' teams in the first gameweek (and the corresponding large points totals seen in \ref{fig:points_summary}) is particularly interesting given that this is before they have observed a physical game being played in the actual season. This suggests a similar approach in identifying players based purely upon their historical performance and corresponding value by the more skilled managers.
	
	\section{Discussion}
	
	The increasing popularity of fantasy sports in recent years \cite{fsgaDemographics} enables the quantitative analysis of managers' decision-making through the study of their digital traces.
	The analysis we present in this article considers the game of Fantasy Premier League, which is played by approximately seven million managers. We observe a consistent level of skill among managers in the sense that there exists a considerable correlation between their performance over multiple seasons of the game, in some cases over thirteen years. This result is particularly striking given the stochastic nature of the underlying game upon which it is based. 
	
	Encouraged by these findings, we proceeded to conduct a deeper analysis 
	of the actions taken by a large proportion of the top one million managers from the 2018-19 season of the game. This allowed each decision made by these managers to be analysed using a variety of statistical and graphical tools. 
	We divided the managers into tiers based upon their final position in the game and observed that the managers in the upper echelons consistently outperformed those in lower ones, suggesting that their skill levels are present throughout the season and that their corresponding rank is not dependent on just a small number of events. The skill-based decisions were apparent in all facets of the game, including making good use of transfers, strong financial awareness, and taking advantage of short- and long-term strategic opportunities, such as their choice of captaincy and use of the chips mechanic, see Section~\ref{subsec:chips}.
	
	Arguably the most remarkable observation presented in this article is, however, the emergence of what we coin a \textit{template team} that suggests a form of common collective behaviour occurring between managers. We show that most teams feature a common core group of constituent players at multiple time points in the season. This occurs despite the wide range of possible options for each decision, 
	suggesting that the managers are acting similarly, and particularly so for the top-tier managers as evident by their higher similarity metrics. Such coordinated behaviour by managers suggests an occurrence of the so-called `superstar effect' within fantasy sports just as per their physical equivalent \cite{lucifora2003superstar}, whereby managers independently arrive at a common conclusion on a core group of players who are viewed as crucial to optimal play. A further dimension is added by the fact that the similarity between the teams of better managers is evident even prior to the first event of the season, i.e., they had apparently all made similar (good) decisions even `before a ball was kicked’.
	
	In this article we have focussed on the behaviour of the managers and their decision-making that constitutes their skill levels. The availability of such detailed data offers the potential for further research from a wide range of areas within the field of computational social science. For example, analysis of the complex financial dynamics taking place within the game as a result of the changing player values and the buying/selling decisions made by the managers would be interesting. A second complementary area of research would be the development of algorithms that consider the range of possible options available to managers and give advice on optimizing point returns. 
	Initial analysis has recently been conducted \cite{bhatt2019should} in this area, including the optimal captaincy choice in a given gameweek, and has demonstrated promising results. 
	
	In summary, we believe the results presented here offer an insight into the behaviour of top fantasy sport managers that is indicative of both long-term planning and collective behaviour within their peer group, demonstrating the intrinsic level of skill required to remain among the top positions over several seasons, as observed in this study. We are however aware that the correlations between decisions and corresponding points demonstrated are not perfect, which is in some sense to be expected due to the non-deterministic nature which makes the sport upon which the game is based so interesting to the millions of individuals who enjoy it each week. These outcomes suggest a combination of skill and luck being present in fantasy sport just as in their physical equivalent.

	\section{Methods}
	
	\subsection{Data Collection}
	
	We obtained the data used in this study by accessing approximately 50 million unique URLs through the Fantasy Premier League API. The rankings at the end of the 2018/19 season were obtained through \url{https://fantasy.premierleague.com/api/leagues-classic/{league-id}/standings/} from which we could obtain the entry ID of the top 1 million ranked managers. Using these IDs we then proceeded to obtain the team selections along with other manager quantities for each gameweek of this season that were used in the study through \url{https://fantasy.premierleague.com/api/entry/{entry-id}/event/{GW}/picks/}, we then filtered the data to include only managers for whom we had data for the entirety of the season which consisted of $901,912$ unique managers. The data for individual footballers and their performances were captured via \url{https://fantasy.premierleague.com/api/bootstrap-static/}. Finally, the historical performance data was obtained for 6 million active managers through \url{https://fantasy.premierleague.com/api/entry/{entry-id}/history/}.
	
	\subsection{Calculation of Transfer Quality}
	
	In order to calculate the transfer quality plot shown in Fig.~\ref{fig:decision_plots}(b) we consider the gameweeks in which managers made one transfer and, based upon the value of the player whom they transferred in, determine what fraction of players with the same price or lower the manager could have instead bought for their team. Suppose that in gameweek $G$ the manager transferred out player $x_i$, who had value $q_G(x_i)$, for player $x_j$ who scored $p_G(x_j)$ points in the corresponding gameweek. The calculation involves firstly finding all players the manager could have transferred in, i.e., those with price less than or equal to $q_G(x_i)$ 
	and then determining the fraction $y_G(x_i, x_j)$ of these players who scored more points than the chosen player given the player whom was transferred out. This is calculated by using
	\begin{equation*}
		y_G(x_i, x_j) = \frac{\sum_k \mathbbm{1}\left[q_G(x_{k}) \le q_G(x_i)\right] \cdot \mathbbm{1}\left[p_G(x_{k}) > p_G(x_j)\right]}{\sum_{\ell} \mathbbm{1}\left[q_G(x_{\ell}) \le q_G(x_i)\right]},
	\end{equation*}	
	where $\mathbbm{1}$ represents the indicator function. 
	Using this quantity we proceed to group over the entire season for each tier of manager which allows us to obtain the distribution of the measure itself and finally the probability of making a better transfer which is shown in panel (b) of Fig.~\ref{fig:decision_plots}. 
	
	\subsection{Team Similarity}\label{subsec:jaccard_calc}
	
	With the aim of identifying levels of similarity between the teams of two managers $i$ and $j$ we make use of the Jaccard similarity which is a measure used to describe the overlap between two sets. Denoting by $T^G_i$ the set of players that appeared in the squad of manager $i$ during gameweek $G$ we consider the Jaccard similarity between the teams of managers $i$ and $j$ for gameweek $G$ given by
	\begin{equation*}
		J^G(i,j) = \frac{\left|T^G_i \cap T^G_j\right|}{\left|T^G_i \cup T^G_j\right|},
		\label{eq:jaccard}
	\end{equation*}
	where $|\cdot|$ represents the cardinality of the set. We then proceed to calculate this measure for all $n$ managers which results in a $n \times n$ symmetric matrix $J^G$, the $(i,j)$ element of which is given by the above equation, note that the diagonal elements of this matrix are unity. Calculation of this quantity over all teams is computationally expensive in the sense that one must perform pair-wise comparison of the $n$ teams for each gameweek. As such we instead calculated an estimate of this quantity by taking random samples without replacement of 100 teams from each tier and calculating the measure both over all teams and also within tiers for each gameweek. We repeat this calculation 10,000 times and the average results are those used in the main text and \ref{sm:clusters}.

	\subsection{Cluster Identification of Player Ownership}
	
	As described in the main text, the calculation of clusters within which groups of players co-appear involves taking advantage of the underlying network structure of all sets of teams. The adjacency matrix describing this network is defined by the matrix $A_{ij}^G$ that has entry $(i,j)$ equal to the number of teams within which player $i$ and $j$ co-appear in gameweek $G$. Note that the diagonal entries of this matrix describe the number of teams in which a given player appears gameweek $G$. Using this matrix we identify the clusters via a hierarchical clustering approach, with $k = 4$ clusters determined via analysing the within-cluster sum of squared errors of $k$-means for each cluster using the elbow method as shown in \ref{fig:WSS}. 
	
%
%
%
	
	\begin{acknowledgements}
		Helpful discussions with Kevin Burke, James Fannon, Peter Grinrod, Stephen Kinsella, Renaud Lambiotte, and Sean McBrearty are gratefully acknowledged. This work was supported by Science Foundation Ireland grant numbers 16/IA/4470, 16/RC/3918, 12/RC/2289 P2 and 18/CRT/6049), co-funded by the European Regional Development Fund.~(J.D.O'B and J.P.G). We acknowledge the DJEI/DES/SFI/HEA Irish Centre for High-End Computing (ICHEC) for the provision of computational facilities and support. The funders had no role in study design, data collection, and analysis, decision to publish, or preparation of
		the manuscript.
	\end{acknowledgements}
	
	\bibliography{fpl_bib}
	
	\onecolumngrid
	\renewcommand{\figurename}{}
	\renewcommand{\tablename}{}
	\renewcommand{\thefigure}{Supplementary Figure \arabic{figure}}
	\renewcommand{\thetable}{Supplementary Table \arabic{table}}
	\setcounter{figure}{0}
	\setcounter{section}{0}
	\setcounter{subsection}{0}
	\renewcommand{\theequation}{S\arabic{equation}}
	\addtocontents{toc}{\protect\setcounter{tocdepth}{0}} 
	\renewcommand{\thesection}{Supplementary Note \Roman{section}}
	\clearpage
	\begin{center}
		\textbf{\large Supplemental Materials}
	\end{center}
	\setcounter{page}{1}
	\section{\NoCaseChange{Summary of Rules of Fantasy Premier League}}\label{app:rules}
	
	The decisions made by the managers of Fantasy Premier League are governed by a stringent set of rules \cite{fplRules}. 
	The initial restrictions of the game is that the manager must select a squad of 15 players consisting of two goalkeepers, five defenders, five midfielders, and three forwards. The total value of these players may not exceed \pounds100M and a further restriction is that no more than three players from one club may appear in a given squad. Each week the manager must then select a starting 11 players which must include one goalkeeper and a minimum of three defenders, three midfielders, and one forward, this restriction is known as the formation criterion. These players are the ones whose performance contributes to the managers' points total. The remaining players feature on the `bench' and are ordered by the manager such that if one of their starting players does not appear the first placed bench players' points are given to the manager (provided the formation criterion remains satisfied by said first choice bench player). 
	
	Players selected by a manager are rewarded points based upon their statistical performance during the physical matches they compete in. Points for specific actions vary by the players' position, for example, a defender receives more points for scoring a goal than a forward due to the relative rarity of such an event. In \ref{tab:points} we show the points per position for each of the possible actions the players may be rewarded/penalised for. For a goalkeeper or defender to be classified as keeping a clean sheet they must have played at least 60 minutes, excluding stoppage time. For example, if a defender was substituted in the physical game with the score at 0-0 in the 63rd minute, and their team proceeded to concede a goal then the defender in question would receive clean sheet points but the remaining players would not. Also, in the case of a goal scored directly from a set piece i.e., a free-kick or penalty, the player who was fouled in the awarding of the set piece receives the assist. 
	\begin{table*}[b]
		\centering
		\caption{Points awarded to players based on action by position.}
		\begin{tabular}{lcccc}
			\toprule
			\textbf{Action} & \textbf{Goalkeeper} & \textbf{Defender} & \textbf{Midfielder} & \textbf{Forward} \\
			\cmidrule(lr){2-2} \cmidrule(lr){3-3} \cmidrule(lr){4-4} \cmidrule(lr){5-5} 
			For playing up to 60 minutes & 1 & 1 & 1 & 1 \\
			For playing 60 minutes or more &2 & 2 & 2 & 2 \\
			For each goal scored & 6 & 6 & 5 & 4 \\
			For each assist & 3 & 3 & 3 & 3 \\
			For keeping a clean sheet & 4 & 4 & 1 & - \\
			For every 3 saves made & 1 & - & - & - \\
			For each penalty saved & 5 & - & - & - \\
			For each penalty missed & -2 & -2 & -2 & -2 \\
			For every two goals conceded & -1 & -1 & - & - \\
			For each yellow card & -1 & -1 & -1 & -1 \\
			For each red card & -3 & -3 & -3 & -3 \\
			For each own goal & -2 & -2 & -2 & -2 \\
			\cline{1-5} 
		\end{tabular}
		\label{tab:points}
	\end{table*}
	
	\begin{figure*}[t]
		\centering
		\includegraphics[width = \textwidth]{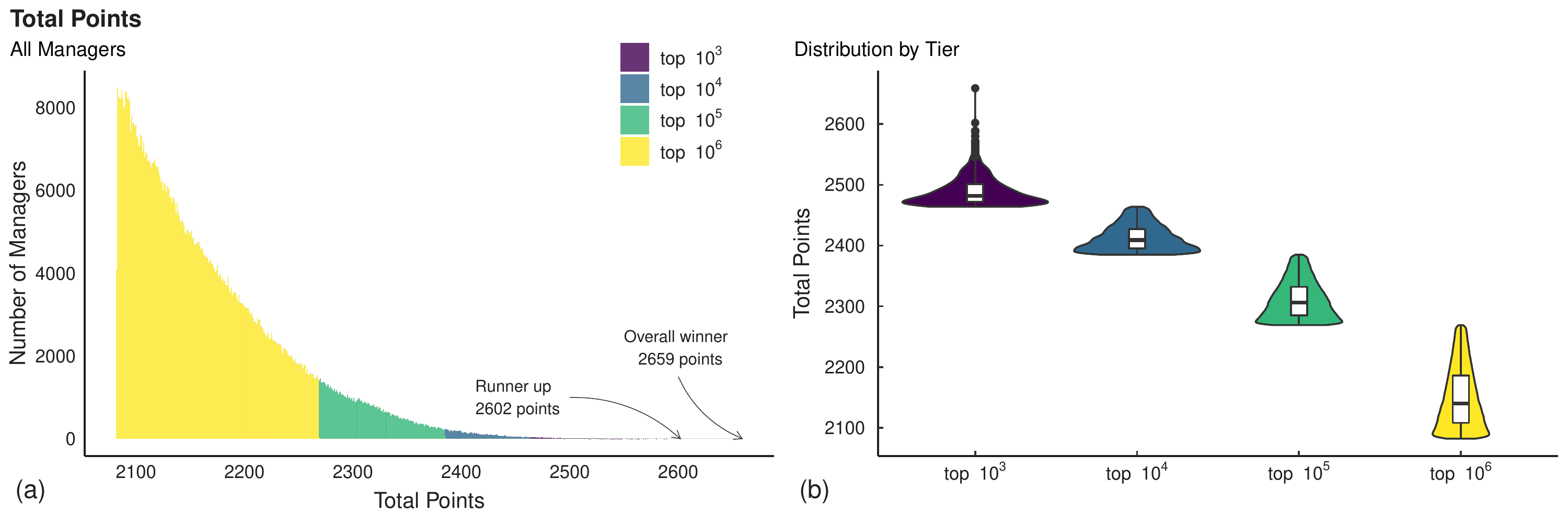}
		\caption{\textbf{Summary of points obtained by managers in the 2018/19 season.}
			(a) The number of managers that obtained each points total where the bins are by tier. The overall winner (2659 points) and second place manager (2602) are highlighted. (b) Distribution of points totals earned by each tier.} 
		\label{fig:points_summary}
	\end{figure*}
	
	\begin{table*}[t]
		\centering
		\caption{Average points and standard deviation of points earned for each tier in each gameweek.}
		\begin{tabular}{p{0.5cm}*{8}{P{1.5cm}}}
			& \multicolumn{2}{c}{$10^3$} & \multicolumn{2}{c}{$10^4$} & \multicolumn{2}{c}{$10^5$} & \multicolumn{2}{c}{$10^6$} \\
			GW & Mean & SD & Mean & SD & Mean & SD & Mean & SD \\
			\cmidrule(lr){2-3} \cmidrule(lr){4-5} \cmidrule(lr){6-7} \cmidrule(lr){8-9}
			1 & 88.16 & 11.73 & 84.97 & 12.65 & 76.84 & 14.62 & 63.17 & 15.06 \\ 
			2 & 85.44 & 11.09 & 83.58 & 12.18 & 78.15 & 14.57 & 68.49 & 16.28 \\ 
			3 & 51.12 & 8.52 & 50.04 & 8.66 & 49.42 & 9.73 & 49.81 & 11.12 \\ 
			4 & 51.85 & 7.74 & 50.80 & 7.99 & 49.70 & 8.97 & 47.00 & 10.23 \\ 
			5 & 71.64 & 14.77 & 68.42 & 14.67 & 64.03 & 14.99 & 54.69 & 15.51 \\ 
			6 & 62.97 & 8.18 & 61.84 & 8.27 & 59.51 & 8.89 & 55.40 & 9.52 \\ 
			7 & 62.80 & 10.36 & 60.83 & 10.22 & 58.13 & 10.61 & 53.96 & 11.33 \\ 
			8 & 70.82 & 13.01 & 66.73 & 13.06 & 62.52 & 13.41 & 56.79 & 13.86 \\ 
			9 & 45.70 & 7.82 & 44.79 & 8.39 & 43.35 & 9.12 & 42.78 & 10.28 \\ 
			10 & 75.62 & 11.34 & 73.70 & 12.02 & 70.56 & 13.93 & 65.21 & 15.70 \\ 
			11 & 70.94 & 12.03 & 68.71 & 12.62 & 65.65 & 13.31 & 59.63 & 14.15 \\ 
			12 & 62.71 & 7.19 & 60.57 & 7.87 & 57.62 & 8.58 & 53.44 & 8.97 \\ 
			13 & 48.36 & 10.92 & 48.18 & 11.27 & 49.19 & 11.56 & 50.79 & 11.99 \\ 
			14 & 59.28 & 7.91 & 58.17 & 8.35 & 56.20 & 9.07 & 53.60 & 10.26 \\ 
			15 & 60.74 & 11.77 & 58.48 & 11.72 & 55.27 & 11.95 & 51.20 & 12.16 \\ 
			16 & 72.55 & 16.74 & 66.80 & 16.36 & 62.70 & 16.04 & 58.82 & 16.41 \\ 
			17 & 55.06 & 9.12 & 53.83 & 9.73 & 51.50 & 10.43 & 47.06 & 11.19 \\ 
			18 & 59.70 & 13.33 & 58.41 & 13.37 & 57.85 & 14.13 & 57.44 & 15.02 \\ 
			19 & 78.08 & 12.44 & 75.98 & 12.60 & 73.45 & 12.63 & 68.13 & 12.63 \\ 
			20 & 61.98 & 11.63 & 58.77 & 11.75 & 55.47 & 12.43 & 52.15 & 12.95 \\ 
			21 & 56.22 & 11.29 & 56.24 & 11.29 & 56.49 & 11.41 & 56.17 & 11.61 \\ 
			22 & 67.43 & 9.04 & 65.45 & 9.47 & 61.50 & 9.95 & 55.73 & 10.31 \\ 
			23 & 68.75 & 8.69 & 67.69 & 9.50 & 65.51 & 10.69 & 60.86 & 12.14 \\ 
			24 & 47.98 & 9.15 & 46.86 & 9.42 & 46.36 & 10.19 & 45.84 & 11.15 \\ 
			25 & 86.11 & 17.93 & 81.53 & 18.19 & 78.68 & 18.62 & 73.11 & 18.48 \\ 
			26 & 74.72 & 10.57 & 72.30 & 11.05 & 69.19 & 11.97 & 64.48 & 13.46 \\ 
			27 & 42.87 & 9.86 & 41.67 & 9.63 & 40.97 & 9.97 & 39.32 & 10.57 \\ 
			28 & 59.38 & 10.03 & 58.44 & 10.65 & 57.36 & 11.60 & 56.91 & 12.90 \\ 
			29 & 47.24 & 8.15 & 45.75 & 8.51 & 45.15 & 9.10 & 43.86 & 9.49 \\ 
			30 & 63.76 & 16.04 & 60.68 & 16.08 & 57.02 & 15.66 & 53.62 & 14.90 \\ 
			31 & 35.11 & 10.41 & 35.26 & 10.96 & 36.44 & 11.76 & 34.57 & 13.45 \\ 
			32 & 98.99 & 12.11 & 96.09 & 12.93 & 90.78 & 14.46 & 80.69 & 15.62 \\ 
			33 & 71.58 & 12.06 & 68.07 & 13.36 & 60.56 & 16.10 & 46.85 & 17.59 \\ 
			34 & 50.87 & 11.80 & 51.96 & 12.95 & 54.25 & 14.33 & 58.19 & 15.21 \\ 
			35 & 106.96 & 17.31 & 100.75 & 18.72 & 89.80 & 19.20 & 76.07 & 15.46 \\ 
			36 & 100.31 & 16.82 & 96.12 & 16.73 & 90.28 & 17.11 & 82.24 & 17.22 \\ 
			37 & 49.19 & 10.50 & 48.02 & 11.16 & 48.67 & 12.35 & 50.77 & 13.20 \\ 
			38 & 66.81 & 11.53 & 66.20 & 11.96 & 64.68 & 12.63 & 61.25 & 13.74 \\ 
			\bottomrule
		\end{tabular}
		\label{tab:manager_points}
	\end{table*}
	
	Managers may proceed to make changes to their team between gameweeks which involves \textit{transferring} a player from their team for another with the same position. Each week a manager is entitled to one such transfer known as a \textit{free transfer}. Additional transfers may be made but at a cost of four points each from their points total. If the manager does not make use of their free transfer the following gameweek they may then make two free transfers however it is not possible to accumulate more than two free transfers. The aforementioned restrictions regarding positions, clubs, and value must be satisfied for each transfer.
	
\begin{figure*}[b]
		\centering
		\includegraphics[width = \textwidth]{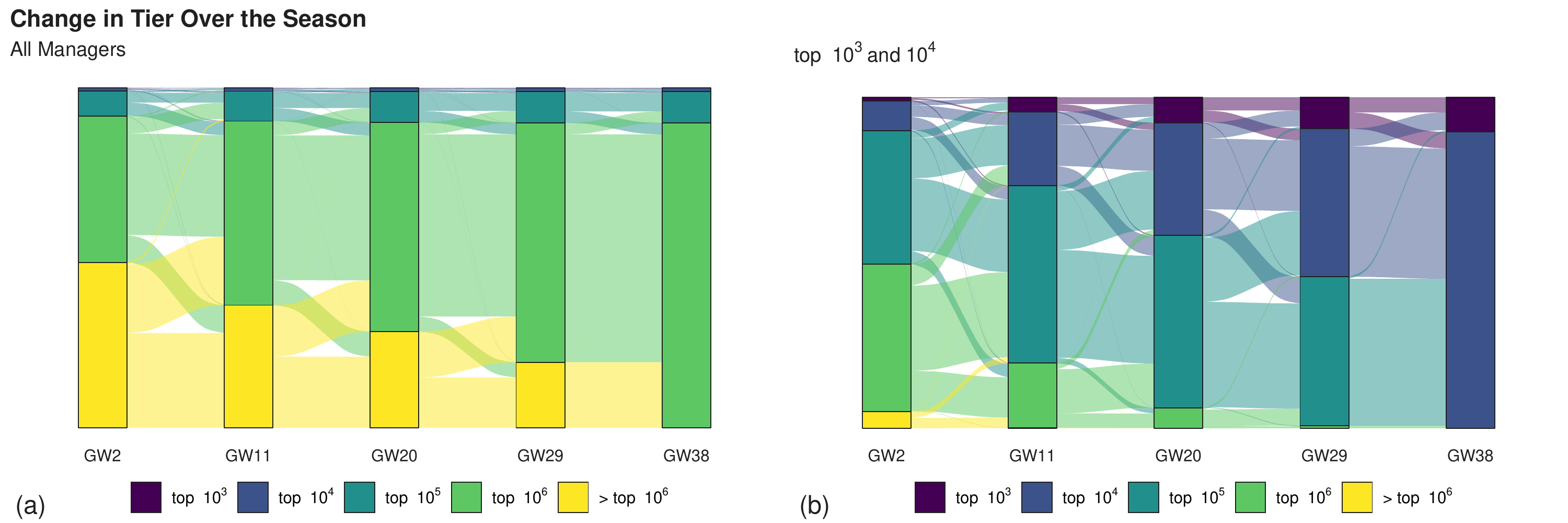}
		\caption{\textbf{Alluvial graph describing the flow of manager rank at multiple
				time points during the season.} (a) Change in tier over all managers in the dataset. (b) The same analysis but on those who finished in the top $10^3$ and $10^4$ tiers.} 
		\label{fig:alluvial_plots}
	\end{figure*}

	In terms of the points amassed by the managers observed in the study we show the points totals and the number of manager who obtained them in panel (a) of \ref{fig:points_summary}. We firstly comment on the skewness of the distribution with the number of managers obtaining a certain number of points decreasing quickly as the points becomes larger. The large gap between the points obtained by the overall winner (2659) and those of second place (2602) is also interesting. The distribution of points within each tier is shown in \ref{fig:points_summary}(b), we see the same skewness being present in each of these distributions, and also the presence of outliers among the top ranked positions. Summary statistics for each tier and also all managers are given in \ref{tab:manager_points} and \ref{table:points_summary}. To view how the ranks of managers change over the course of the season we show the flow of manager position in \ref{fig:alluvial_plots}. All managers are considered in panel (a) while those who finished in the top $10^4$ positions are shown in (b). We see the competition for the top positions by
	the fact that no manager who is outside the top $10^6$ ranks at gameweek 20 finishes within the top $10^6$ tier.
	
	\begin{table*}
		\centering
		\caption{Summary statistics of the points obtained by the managers in the dataset. Both over all managers and the tiers used in this study.}
		\begin{tabular}{p{1.5cm}*{5}{P{2.5cm}}}
			\toprule
			& \multicolumn{5}{c}{Tier} \\
			\cmidrule[\heavyrulewidth]{2-6}
			& Everyone & $10^3$ & $10^4$ & $10^5$ & $10^6$ \\
			\cmidrule(lr){2-2} \cmidrule(lr){3-3} \cmidrule(lr){4-4} \cmidrule(lr){5-5} \cmidrule(lr){6-6}
			n &901912&1000&8493&83897&808522 \\
			Max &2659&2659&2464&2385&2269 \\
			Min &2082&2464&2385&2269&2082 \\
			Mean &2167.91&2489.82&2412.68&2310.82&2150.11 \\
			Median &2150&2482&2409&2306&2140 \\
			Std. dev &71.72&24.21&20.68&30.44&49.42 \\
			IQR &96&29&32&47&78 \\
			\bottomrule
		\end{tabular}
		\label{table:points_summary}
	\end{table*}
	
	\section{\NoCaseChange{Historical Correlations}}
	
	To obtain the historical data we considered managers taking part in the 2019/20 season of the game for which we found $\approx6$ million managers, of whom $\approx3.8$ million had taken part in a previous season. We then determined their historical performance in terms of both points earned and overall rank for each season in which they partook in the game. These quantities are the only two available at the historical level, unlike the gameweek level of resolution studied in the main text. We then proceeded to identify the number of managers for whom we have data in each pairwise combination of seasons, shown in the lower elements of \ref{table:correlations}, and calculated Pearson correlations between their points totals in each case represented by the upper elements of the same table. These correlations are visualised in Fig.~\ref{fig:history_plots} of the main text. The two single blocks in the figure represent the winner (who had played in six previous seasons) and the runner-up (three previous seasons).
	
\begin{figure}[h!]
		\centering
		\includegraphics[width = 0.75\textwidth]{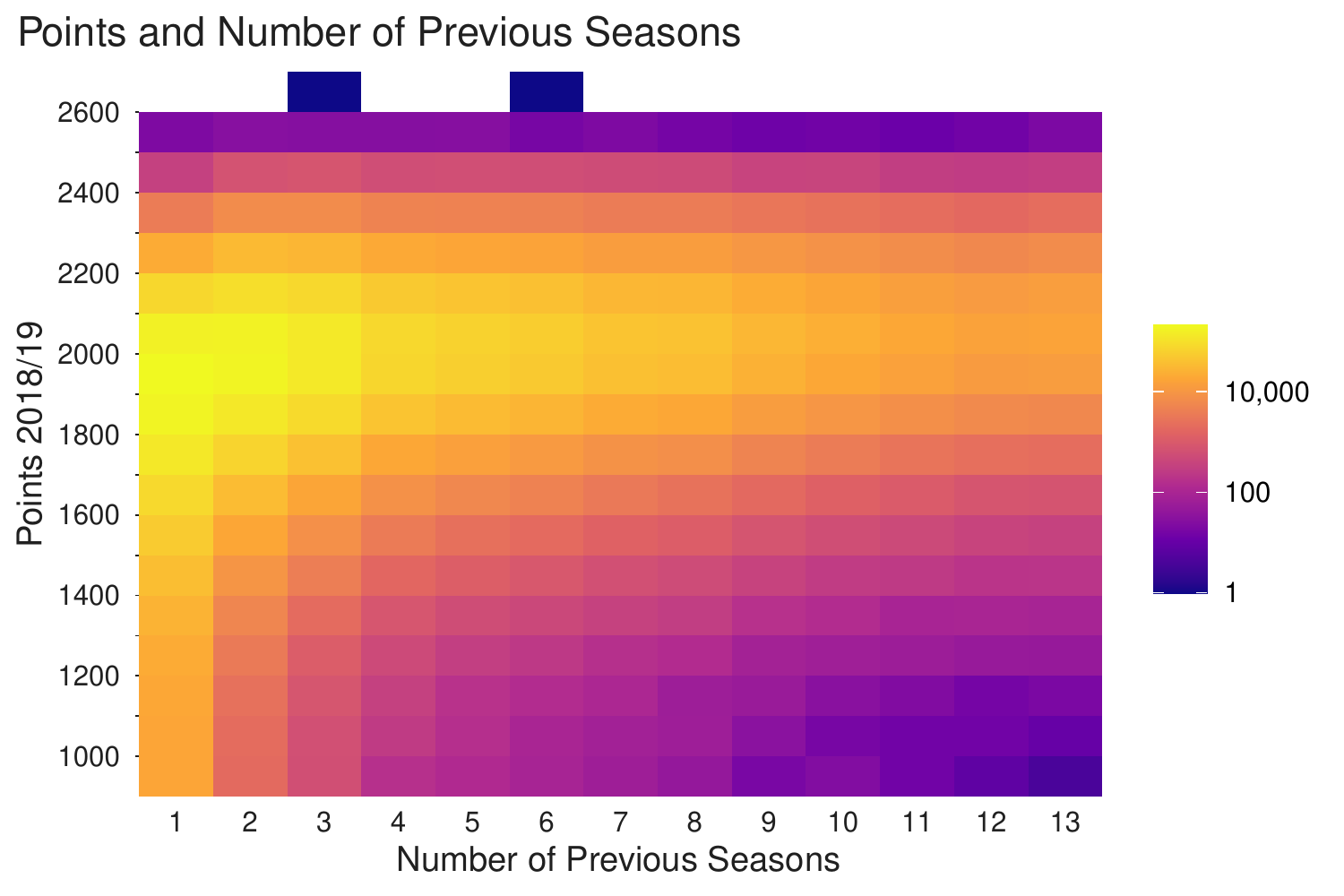}
		\caption{\textbf{Manager points in the 2018/19 season versus the number of previous seasons which they had registered for.} The bins, each of which cover a 100 point range, are coloured by the number of managers in each, note the logarithmic scale. We comment on the small number present in the bottom right corner (in comparison to the top right corner), which indicates that players who have played for multiple years did not tend to perform poorly.} 
		\label{fig:points_history}
	\end{figure}

	We also consider the relationship between the number of previous seasons that the managers took part in and their points totals in the 2018/19 seasons as shown in \ref{fig:points_history}. We comment on the small number of managers present in the bottom right corner of the plot, suggesting that managers who have played for multiple years are less likely to have obtained a lower points total, in line with the correlation results obtained in \ref{table:correlations}. This positive relationship is also evident when one fits a linear regression to the data which suggests each additional year's experience is worth, on average, an additional 22 points.
	
	\begin{table*}[t]
		\centering
		\caption{Correlation between a managers' historical performance. The lower triangular elements of the table represent the number of managers who were present in both of the seasons, such that the diagonal elements describe the number of managers for each season for whom we could obtain data. Upper elements of the table represent the pairwise Pearson correlation coefficient between the points obtained by the manager in the two seasons.}
		\setlength{\tabcolsep}{3pt}
		\resizebox{\textwidth}{!}{\begin{tabular}{lccccccccccccc}	\toprule
				\textbf{Season} & 2018-19 & 2017-18 & 2016-17 & 2015-16 & 2014-15 & 2013-14 & 2012-13 & 2011-12 & 2010-11 & 2009-10 & 2008-09 & 2007-08 & 2006-07 \\ 
				\cmidrule{2-14}
				2018-19 & 3,810,484 & 0.42 & 0.36 & 0.30 & 0.30 & 0.26 & 0.25 & 0.21 & 0.20 & 0.20 & 0.18 & 0.13 & 0.14 \\ 
				2017-18 & 2,577,419 & 2,951,332 & 0.43 & 0.36 & 0.34 & 0.31 & 0.29 & 0.25 & 0.24 & 0.23 & 0.21 & 0.15 & 0.15 \\ 
				2016-17 & 1,854,775 & 1,876,837 & 2,114,618 & 0.43 & 0.41 & 0.33 & 0.31 & 0.28 & 0.25 & 0.25 & 0.21 & 0.15 & 0.15 \\ 
				2015-16 & 1,407,603 & 1,410,074 & 1,381,427 & 1,549,899 & 0.46 & 0.36 & 0.31 & 0.30 & 0.27 & 0.24 & 0.20 & 0.15 & 0.15 \\ 
				2014-15 & 1,163,917 & 1,171,594 & 1,134,400 & 1,153,343 & 1,279,695 & 0.47 & 0.40 & 0.36 & 0.32 & 0.31 & 0.25 & 0.18 & 0.18 \\ 
				2013-14 & 963,170 & 969,670 & 933,216 & 939,339 & 946,672 & 1,058,033 & 0.45 & 0.35 & 0.30 & 0.28 & 0.25 & 0.19 & 0.19 \\ 
				2012-13 & 754,449 & 758,702 & 731,609 & 734,445 & 733,617 & 741,797 & 823,537 & 0.46 & 0.37 & 0.35 & 0.29 & 0.23 & 0.22 \\ 
				2011-12 & 635,833 & 638,131 & 612,120 & 611,371 & 606,032 & 604,716 & 604,303 & 692,553 & 0.45 & 0.40 & 0.29 & 0.25 & 0.23 \\ 
				2010-11 & 420,544 & 421,508 & 409,055 & 411,163 & 409,505 & 409,952 & 412,101 & 440,594 & 453,150 & 0.44 & 0.34 & 0.27 & 0.26 \\ 
				2009-10 & 315,106 & 315,569 & 306,943 & 308,726 & 307,495 & 308,039 & 309,390 & 329,926 & 313,801 & 338,437 & 0.44 & 0.34 & 0.33 \\ 
				2008-09 & 223,217 & 223,374 & 217,829 & 219,255 & 218,378 & 218,595 & 219,578 & 233,607 & 221,887 & 223,383 & 239,153 & 0.37 & 0.35 \\ 
				2007-08 & 153,640 & 153,615 & 149,788 & 150,831 & 150,241 & 150,298 & 150,976 & 160,669 & 151,297 & 150,545 & 150,766 & 164,385 & 0.38 \\ 
				2006-07 & 89,260 & 89,182 & 87,197 & 87,819 & 87,455 & 87,504 & 87,994 & 93,241 & 88,130 & 87,250 & 86,672 & 87,762 & 95,231 \\ 
				\hline
		\end{tabular}}
		\label{table:correlations}
	\end{table*}
	
	\section{\NoCaseChange{Financial Analysis}}\label{sm:finance}
	
\begin{figure*}[h!]
		\centering
		\includegraphics[width = \textwidth]{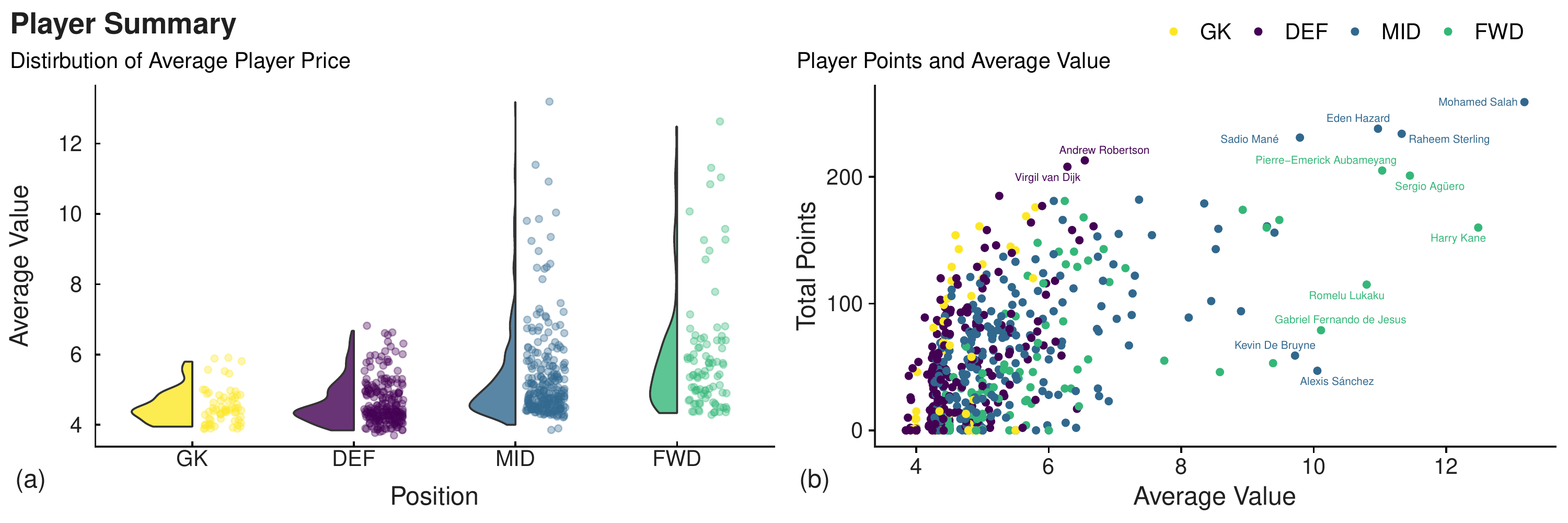}
		\caption{\textbf{The value of player values and the corresponding points earned.} (a) Distribution of average player price over the course of the season factored by the player position, we comment on the skewed nature of the distribution particularly for midfielders and forwards. (b) The same average value is shown versus the corresponding points earned over the season by said players, which shows the largest points totals being provided by generally the higher priced players. The identities of some players with higher prices and points totals are also shown.} 
		\label{fig:player_points}
	\end{figure*}

	Here we briefly consider implications of the financial aspects of the game. Firstly, the price of players themselves demonstrate some interesting dynamics. We remind the reader that initially the price of each player is set by the developers of the game and this price subsequently fluctuates over the season depending on the supply and demand of managers transferring the player in and out of their teams. A summary of the distribution of the average value over the season for each of the $\approx$600 players coloured by their corresponding position is provided in panel (a) of \ref{fig:player_points}. We comment on the skewed distribution of points obtained in all cases but in particular for midfielders and forwards. The corresponding points earned over the season by each of the players versus their average price is shown in panel (b). We see that there is, in general, a positive relationship between the price of a player and their corresponding points totals. As per their prices, we see a handful of midfielders and forwards who earn the most points.
	
\begin{figure}[h!]
		\centering
		\includegraphics[width = \textwidth]{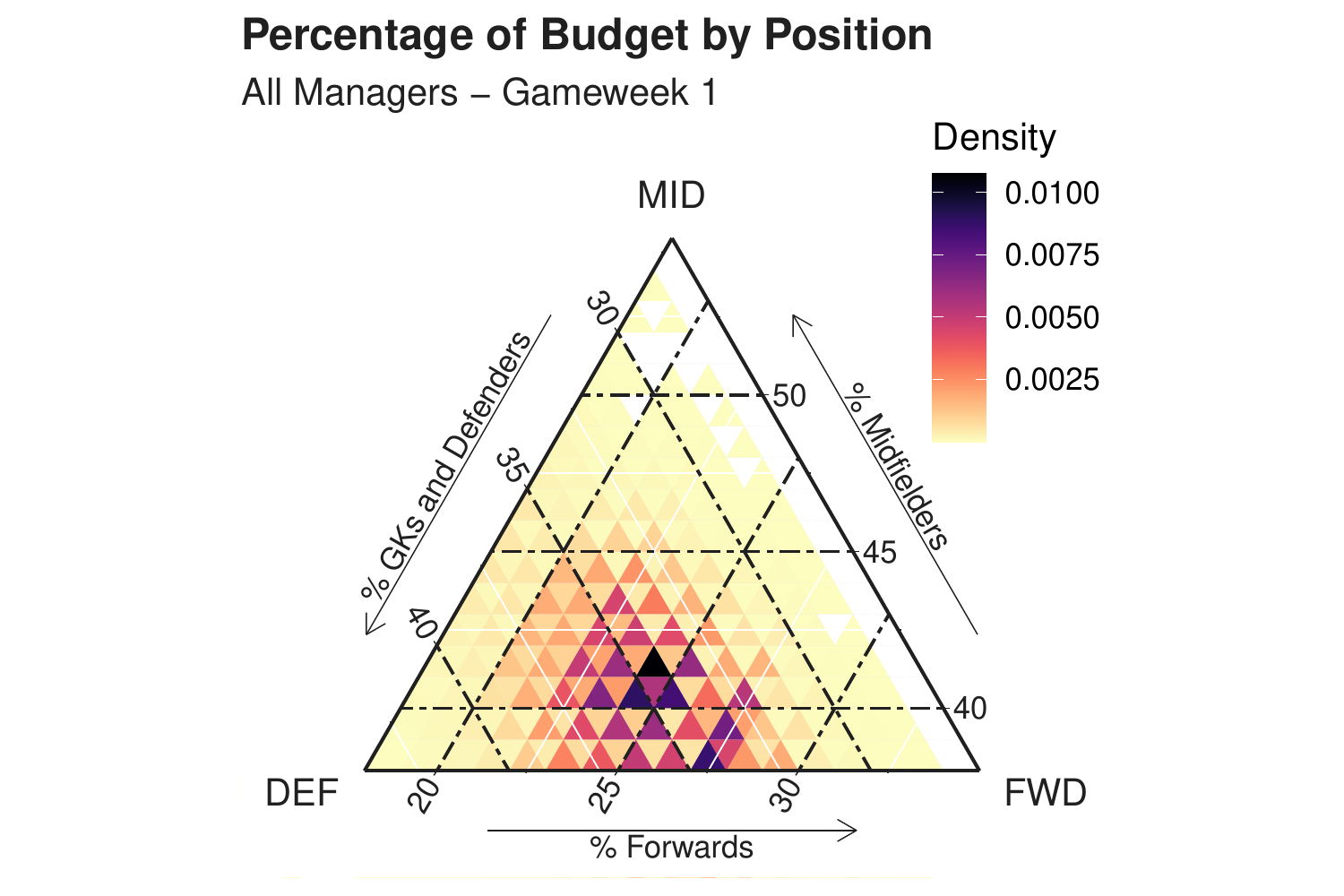}
		\caption{\textbf{Ternary diagram demonstrating the make-up of managers' squads in gameweek one.} The combination of proportions spent in each position (where DEF represents both goalkeepers and defenders) is shown, where the colour in each bin represents the fraction of managers who used a given combination of proportions.} 
		\label{fig:ternary}
	\end{figure}

\begin{figure}[h!]
		\centering
		\includegraphics[width =0.75\textwidth]{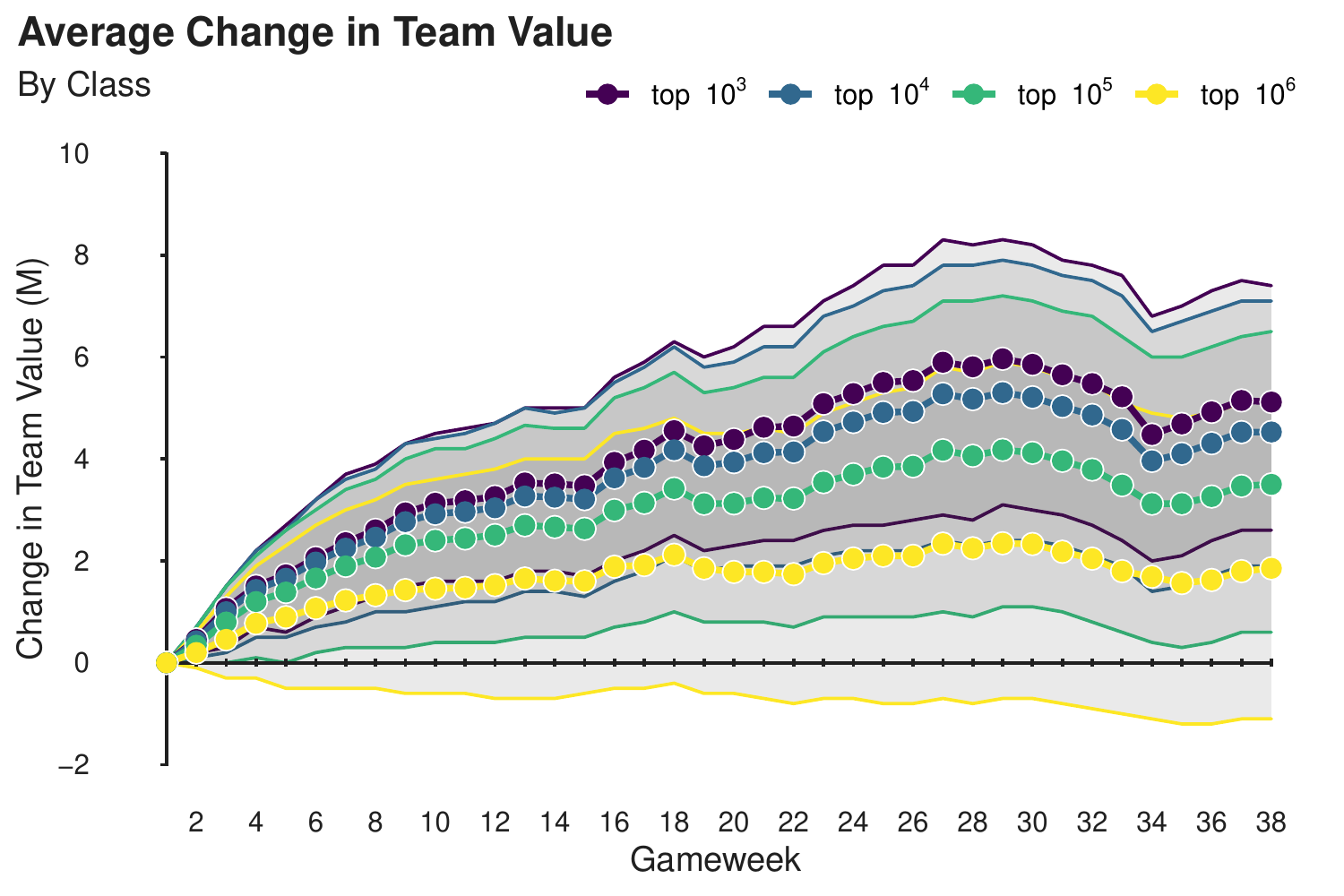}
		\caption{\textbf{Average team value along with 95 percentiles for each tier over the season.} We comment on the general upward trend, but observe the higher-placed managers having larger team values throughout the season.} 
		\label{fig:tv_class}
	\end{figure}

	The make-up of the managers' squad consists of two goalkeepers, five defenders, five midfielders, and three forwards. \ref{fig:ternary} demonstrates the proportion of budget spend in these three positions (we have grouped goalkeepers and defenders together) by all managers at GW 1. We observe some remarkable variation in where the budget is spent, with some managers spending over half of their budget on midfield players despite of them only accounting for a third of their squad. Due to the price of players fluctuating throughout the season, like an investor holding a varying stock, the managers' overall team value changes. \ref{fig:tv_class} shows the average team value of the managers by tier over each gameweek of the season along with corresponding 95\% intervals of the distribution. As per the main text we fit a linear regression to the total points obtained by all managers as a function of their team value each gameweek, with results shown in \ref{tab:regression_coefficients}.
	
	\begin{table*}[ht]
		\centering
		\caption{Regression coefficients for final points as a function of each 
			additional million pounds in team value at each gameweek over all managers.}
		\label{tab:regression_coefficients}
		\vspace{-3mm}
		\begin{tabular}{l @{\extracolsep{15pt}} cccc}
			GW & Intercept & Co-efficient & $p$ & $R^2$ \\ 
			\cmidrule(lr){2-2}
			\cmidrule(lr){3-3}
			\cmidrule(lr){4-4}
			\cmidrule(lr){5-5}
			1 & 2167.907 & --- & --- & $ < 10^{-6} $ \\ 
			2 & 2143.709 & 110.991 & $ < 10^{-6} $ & 0.092 \\ 
			3 & 2141.321 & 53.410 & $ < 10^{-6} $ & 0.107 \\ 
			4 & 2136.398 & 38.339 & $ < 10^{-6} $ & 0.097 \\ 
			5 & 2141.617 & 27.676 & $ < 10^{-6} $ & 0.082 \\ 
			6 & 2138.070 & 26.221 & $ < 10^{-6} $ & 0.094 \\ 
			7 & 2135.709 & 24.810 & $ < 10^{-6} $ & 0.102 \\ 
			8 & 2133.034 & 24.766 & $ < 10^{-6} $ & 0.112 \\ 
			9 & 2133.060 & 22.865 & $ < 10^{-6} $ & 0.124 \\ 
			10 & 2131.917 & 23.098 & $ < 10^{-6} $ & 0.133 \\ 
			11 & 2131.612 & 23.023 & $ < 10^{-6} $ & 0.136 \\ 
			12 & 2132.525 & 21.688 & $ < 10^{-6} $ & 0.130 \\ 
			13 & 2130.013 & 21.430 & $ < 10^{-6} $ & 0.137 \\ 
			14 & 2129.650 & 22.168 & $ < 10^{-6} $ & 0.143 \\ 
			15 & 2130.758 & 21.695 & $ < 10^{-6} $ & 0.138 \\ 
			16 & 2127.116 & 20.324 & $ < 10^{-6} $ & 0.139 \\ 
			17 & 2124.623 & 21.092 & $ < 10^{-6} $ & 0.158 \\ 
			18 & 2120.340 & 21.016 & $ < 10^{-6} $ & 0.167 \\ 
			19 & 2124.464 & 21.789 & $ < 10^{-6} $ & 0.169 \\ 
			20 & 2124.513 & 22.473 & $ < 10^{-6} $ & 0.185 \\ 
			21 & 2124.235 & 22.474 & $ < 10^{-6} $ & 0.198 \\ 
			22 & 2125.238 & 22.415 & $ < 10^{-6} $ & 0.202 \\ 
			23 & 2122.807 & 21.153 & $ < 10^{-6} $ & 0.203 \\ 
			24 & 2122.490 & 20.379 & $ < 10^{-6} $ & 0.204 \\ 
			25 & 2122.837 & 19.661 & $ < 10^{-6} $ & 0.207 \\ 
			26 & 2124.070 & 19.118 & $ < 10^{-6} $ & 0.204 \\ 
			27 & 2122.017 & 18.063 & $ < 10^{-6} $ & 0.202 \\ 
			28 & 2124.113 & 17.892 & $ < 10^{-6} $ & 0.198 \\ 
			29 & 2122.244 & 17.922 & $ < 10^{-6} $ & 0.201 \\ 
			30 & 2122.089 & 18.120 & $ < 10^{-6} $ & 0.198 \\ 
			31 & 2123.304 & 18.771 & $ < 10^{-6} $ & 0.204 \\ 
			32 & 2126.808 & 18.373 & $ < 10^{-6} $ & 0.196 \\ 
			33 & 2130.075 & 19.057 & $ < 10^{-6} $ & 0.195 \\ 
			34 & 2134.159 & 18.271 & $ < 10^{-6} $ & 0.162 \\ 
			35 & 2135.771 & 18.504 & $ < 10^{-6} $ & 0.176 \\ 
			36 & 2133.740 & 18.915 & $ < 10^{-6} $ & 0.188 \\ 
			37 & 2130.639 & 18.793 & $ < 10^{-6} $ & 0.191 \\ 
			38 & 2131.246 & 18.008 & $ < 10^{-6} $ & 0.181 \\ 
			\bottomrule
		\end{tabular}
	\end{table*}

	\section{\NoCaseChange{Team Similarity and Cluster Analysis}}\label{sm:clusters}
	
\begin{figure}[h!]
		\centering
		\includegraphics[width = 0.75\textwidth]{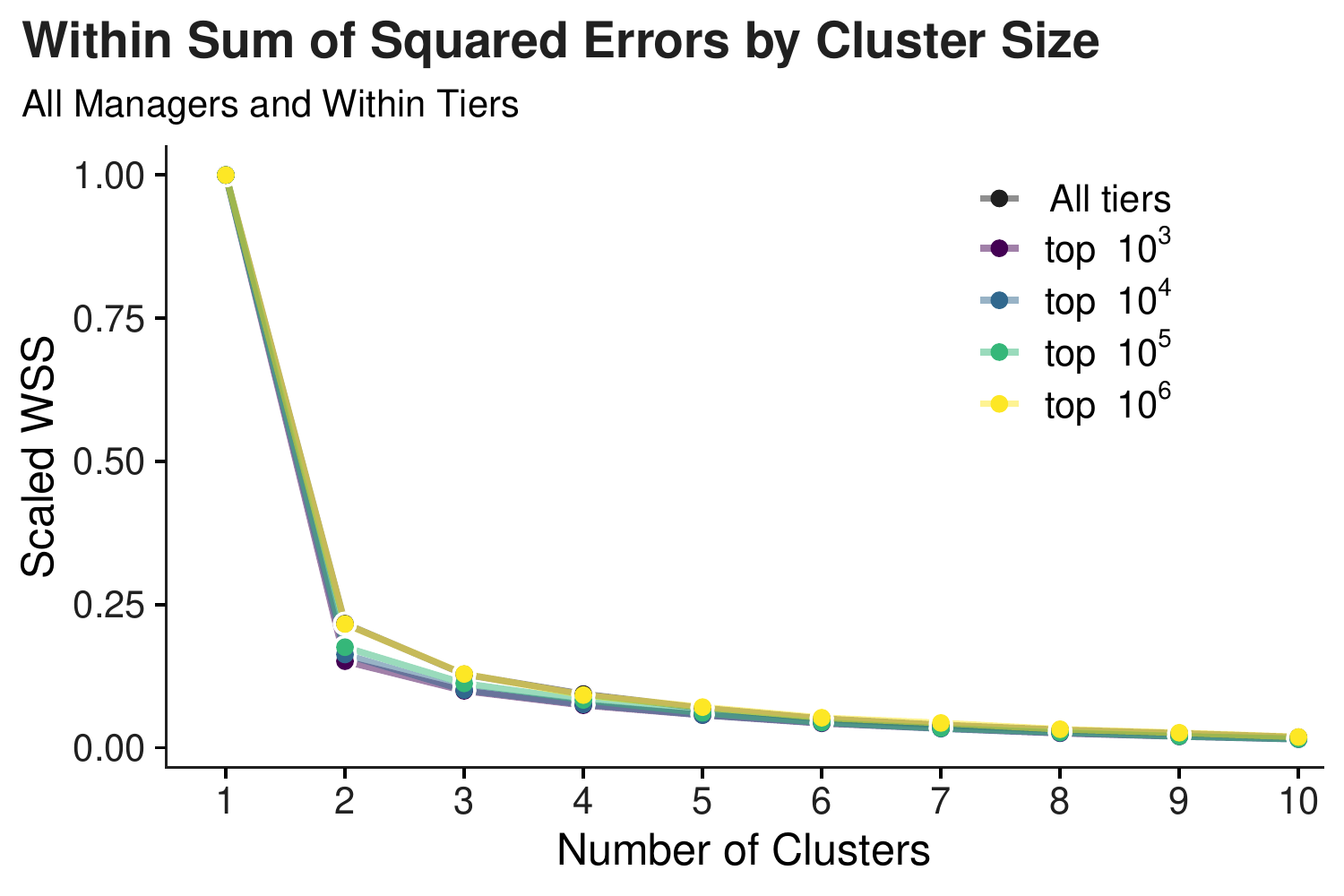}
		\caption{\textbf{Scaled within sum of squared errors for the $k$-means cluster
				analysis} The horizontal axis represents the number of clusters used and the vertical the within sum of squared errors. The measure is calculated for each tier in each of the 38 gameweeks before being rescaled in order to give equal weighting to each gameweek. We note that each tier follows a similar pattern.} 
		\label{fig:WSS}
	\end{figure}

	In this section we show more information regarding the hierarchical clustering analysis described in the main text. \ref{fig:WSS} shows the scaled within sum of squared errors (WSS) for each number of clusters where the error is calculated using $k$-means. We decide upon four clusters as the decrease in errors slows down at this point. In order to give equal weighting to each gameweek we firstly calculate the WSS for each gameweek and rescale these before averaging over these rescaled values over all gameweeks.
	
\begin{figure*}
		\centering
		\includegraphics[width = \textwidth]{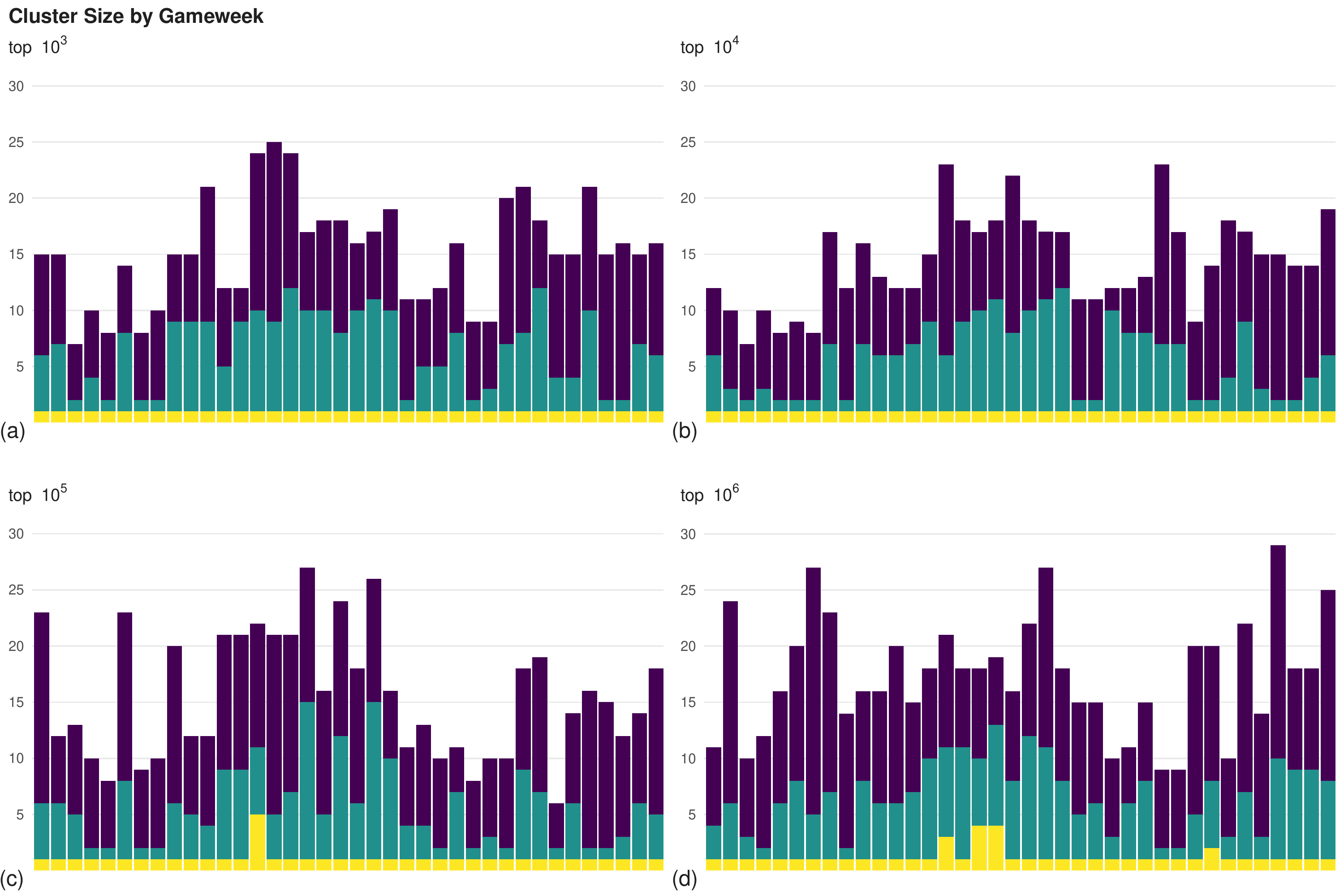}
		\caption{\textbf{Cluster size analysis.} Size of the first three clusters identified by the hierarchical clustering approach described in the main text for each tier. Note all appear to follow a similar pattern.}
		\label{fig:clusters_all}
	\end{figure*}

\begin{figure*}
		\centering
		\includegraphics[width = \textwidth]{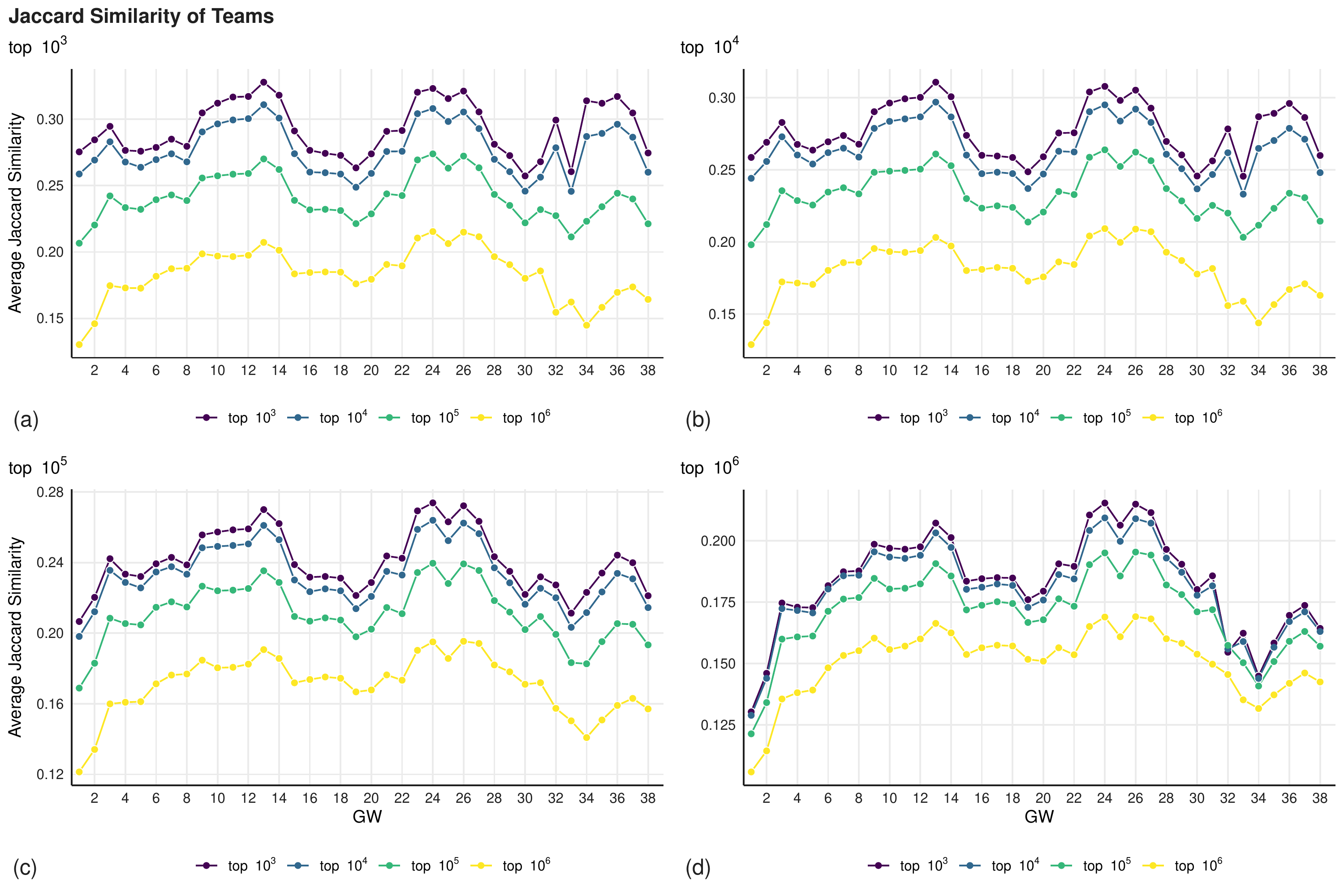}
		\caption{\textbf{Jaccard similarity between managers in each tier with those in other tiers.} Calculation is repeated as described in Sec.~\ref{subsec:jaccard_calc} of the main text. The Jaccard similarity of teams in each tier is compared with all other tiers, where we observe a stronger similarity between those in higher tiers indicating that the better managers are more likely to have a similar structure.} 
		\label{fig:jaccard_all}
	\end{figure*}

	The sizes of these first three clusters for each tier are shown in \ref{fig:clusters_all} and follow a similar pattern to that found for all managers in the main text. The top two tiers are shown in panel (a) and (b), however, do appear to make use of fewer players which may be a function of the smaller number of teams to analyse It may also be further evidence of the higher similarity between the teams in these tiers as suggested in \ref{fig:jaccard_all}, which shows the Jaccard similarity (calculated as in the main text) for each of the tiers versus all other tiers. Finally, for the interested reader we provide the identity of those players who appear in the first cluster when the analysis is performed on all managers in \ref{table:first_cluster}. We see frequent appearance of some higher priced players such as Mohammed Salah and Sergio Agüero throughout the season. However, also interesting is the presence of some extremely inexpensive players, in particular Aaron Wan-Bissaka who was appearing in his debut campaign and was priced at the cheapest level as a \pounds 4M defender but surprisingly made consistent appearances throughout the season, which made him a very attractive option for skilful managers in order to spend more of their budget elsewhere. 
	
	\begin{table*}
		\centering
		\caption{Summary of players who appeared in the first cluster over the course of the season among the different tiers of managers.}
		\resizebox{0.9\textwidth}{!}{\begin{tabular}{c @{\extracolsep{\fill}} ccccc}
				& \multicolumn{5}{c}{Tier} \\
				\cmidrule[\heavyrulewidth]{2-6}
				Gameweek & Everyone & $10^3$ & $10^4$ & $10^5$ & $10^6$ \\
				\cmidrule(lr){1-1} \cmidrule(lr){2-6}
				1 & Mohamed Salah & Sergio Agüero & Sergio Agüero & Mohamed Salah & Mohamed Salah \\ 
				\cmidrule(lr){2-6}
				2 & Mohamed Salah & Sergio Agüero & Sergio Agüero & Sergio Agüero & Mohamed Salah \\
				\cmidrule(lr){2-6} 
				3 & Mohamed Salah & Sergio Agüero & Sergio Agüero & Sergio Agüero & Sergio Agüero \\
				\cmidrule(lr){2-6} 
				4 & Sergio Agüero & Sergio Agüero & Sergio Agüero & Sergio Agüero & Sergio Agüero \\ 
				\cmidrule(lr){2-6}
				5 & Sergio Agüero & Sergio Agüero & Sergio Agüero & Sergio Agüero & Sergio Agüero \\ 
				\cmidrule(lr){2-6}
				6 & Sergio Agüero & Aaron Wan-Bissaka & Aaron Wan-Bissaka & Sergio Agüero & Sergio Agüero \\ 
				\cmidrule(lr){2-6}
				7 & Sergio Agüero & Aaron Wan-Bissaka & Aaron Wan-Bissaka & Sergio Agüero & Sergio Agüero \\ 
				\cmidrule(lr){2-6}
				8 & Eden Hazard & Aaron Wan-Bissaka & Aaron Wan-Bissaka & Aaron Wan-Bissaka & Eden Hazard \\
				\cmidrule(lr){2-6} 
				9 & Eden Hazard & Eden Hazard & Eden Hazard & Eden Hazard & Eden Hazard \\ 
				\cmidrule(lr){2-6}
				10 & Eden Hazard & Aaron Wan-Bissaka & Aaron Wan-Bissaka & Aaron Wan-Bissaka & Eden Hazard \\ 
				\cmidrule(lr){2-6}
				11 & Sergio Agüero & Aaron Wan-Bissaka & Aaron Wan-Bissaka & Sergio Agüero & Sergio Agüero \\ 
				\cmidrule(lr){2-6}
				12 & Sergio Agüero & Mohamed Salah & Aaron Wan-Bissaka & Aaron Wan-Bissaka & Sergio Agüero \\ 
				\cmidrule(lr){2-6}
				13 & Sergio Agüero & Mohamed Salah & Sergio Agüero & Sergio Agüero & Sergio Agüero \\ 
				\cmidrule(lr){2-6}
				\multirow{5}{*}{14} & Sergio Agüero & Aaron Wan-Bissaka & Aaron Wan-Bissaka & Marcos Alonso & Sergio Agüero \\ 
				& --- & --- & --- & Aaron Wan-Bissaka & --- \\ 
				& --- & --- & --- & Andrew Robertson & --- \\ 
				& --- & --- & --- & Richarlison de Andrade & --- \\ 
				& --- & --- & --- & Sergio Agüero & --- \\ 
				\cmidrule(lr){2-6}
				\multirow{4}{*}{15} & Aaron Wan-Bissaka & Aaron Wan-Bissaka & Aaron Wan-Bissaka & Aaron Wan-Bissaka & Aaron Wan-Bissaka \\ 
				& Callum Wilson & --- & --- & --- & Callum Wilson \\ 
				& Marcos Alonso & --- & --- & --- & Marcos Alonso \\ 
				& Richarlison de Andrade & --- & --- & --- & \\ 
				\cmidrule(lr){2-6}
				\multirow{2}{*}{16} & Aaron Wan-Bissaka & Aaron Wan-Bissaka & Aaron Wan-Bissaka & Aaron Wan-Bissaka & Callum Wilson \\ 
				& Callum Wilson & --- & --- & --- & --- \\ 
				\cmidrule(lr){2-6}
				\multirow{4}{*}{17} & Aaron Wan-Bissaka & Aaron Wan-Bissaka & Aaron Wan-Bissaka & Aaron Wan-Bissaka & Aaron Wan-Bissaka \\ 
				& Marcos Alonso & --- & --- & --- & Callum Wilson \\ 
				& --- & --- & --- & --- & Marcos Alonso \\ 
				& --- & --- & --- & --- & Pierre-Emerick Aubameyang \\ 
				\cmidrule(lr){2-6}
				\multirow{4}{*}{18} & Aaron Wan-Bissaka & Aaron Wan-Bissaka & Aaron Wan-Bissaka & Aaron Wan-Bissaka & Aaron Wan-Bissaka \\ 
				& --- & --- & --- & --- & Andrew Robertson \\ 
				& --- & --- & --- & --- & Callum Wilson \\ 
				& --- & --- & --- & --- & Marcos Alonso \\ 
				\cmidrule(lr){2-6}
				19 & Aaron Wan-Bissaka & Aaron Wan-Bissaka & Aaron Wan-Bissaka & Aaron Wan-Bissaka & Aaron Wan-Bissaka \\ 
				\cmidrule(lr){2-6}
				20 & Aaron Wan-Bissaka & Aaron Wan-Bissaka & Aaron Wan-Bissaka & Aaron Wan-Bissaka & Aaron Wan-Bissaka \\ 
				\cmidrule(lr){2-6}
				21 & Aaron Wan-Bissaka & Aaron Wan-Bissaka & Aaron Wan-Bissaka & Aaron Wan-Bissaka & Aaron Wan-Bissaka \\ 
				\cmidrule(lr){2-6}
				22 & Aaron Wan-Bissaka & Aaron Wan-Bissaka & Aaron Wan-Bissaka & Aaron Wan-Bissaka & Aaron Wan-Bissaka \\ 
				\cmidrule(lr){2-6} 
				23 & Mohamed Salah & Mohamed Salah & Mohamed Salah & Mohamed Salah & Mohamed Salah \\ 
				\cmidrule(lr){2-6} 
				24 & Marcus Rashford & Mohamed Salah & Mohamed Salah & Mohamed Salah & Marcus Rashford \\ 
				\cmidrule(lr){2-6} 
				25 & Mohamed Salah & Mohamed Salah & Mohamed Salah & Aaron Wan-Bissaka & Mohamed Salah \\ 
				\cmidrule(lr){2-6} 
				26 & Marcus Rashford & Mohamed Salah & Mohamed Salah & Mohamed Salah & Marcus Rashford \\ 
				\cmidrule(lr){2-6} 
				27 & Paul Pogba & Mohamed Salah & Mohamed Salah & Mohamed Salah & Paul Pogba \\ 
				\cmidrule(lr){2-6} 
				28 & Raúl Jiménez & Mohamed Salah & Mohamed Salah & Paul Pogba & Raúl Jiménez \\ 
				\cmidrule(lr){2-6} 
				29 & Paul Pogba & Raúl Jiménez & Paul Pogba & Paul Pogba & Paul Pogba \\ 
				\cmidrule(lr){2-6} 
				30 & Paul Pogba & Mohamed Salah & Mohamed Salah & Paul Pogba & Paul Pogba \\ 
				\cmidrule(lr){2-6} 
				\multirow{3}{*}{31} & Andrew Robertson & Mohamed Salah & Mohamed Salah & Mohamed Salah & Mohamed Salah \\ 
				& Eden Hazard & --- & --- & --- & Sadio Mané \\ 
				& Sadio Mané & --- & --- & --- & --- \\ 
				\cmidrule(lr){2-6} 
				32 & Paul Pogba & Sergio Agüero & Sergio Agüero & Sergio Agüero & Paul Pogba \\ 
				\cmidrule(lr){2-6} 
				33 & Andrew Robertson & Eden Hazard & Eden Hazard & Eden Hazard & Andrew Robertson \\ 
				\cmidrule(lr){2-6} 
				34 & Raúl Jiménez & Raúl Jiménez & Heung-Min Son & Raúl Jiménez & Raúl Jiménez \\ 
				\cmidrule(lr){2-6}
				35 & Raúl Jiménez & Raúl Jiménez & Raúl Jiménez & Raúl Jiménez & Raúl Jiménez \\ 
				\cmidrule(lr){2-6} 
				36 & Raúl Jiménez & Raúl Jiménez & Raúl Jiménez & Raúl Jiménez & Raúl Jiménez \\ 
				\cmidrule(lr){2-6}
				37 & Raúl Jiménez & Raúl Jiménez & Raúl Jiménez & Raúl Jiménez & Raúl Jiménez \\ 
				\cmidrule(lr){2-6}
				38 & Sadio Mané & Sergio Agüero & Sergio Agüero & Sadio Mané & Sadio Mané \\ 
				\bottomrule
		\end{tabular}}
		\label{table:first_cluster}
	\end{table*}
	
	\section{\NoCaseChange{Chip Usage}}\label{sm:chips}
	
\begin{figure*}[t]
		\centering
		\includegraphics[height = 0.75\textheight]{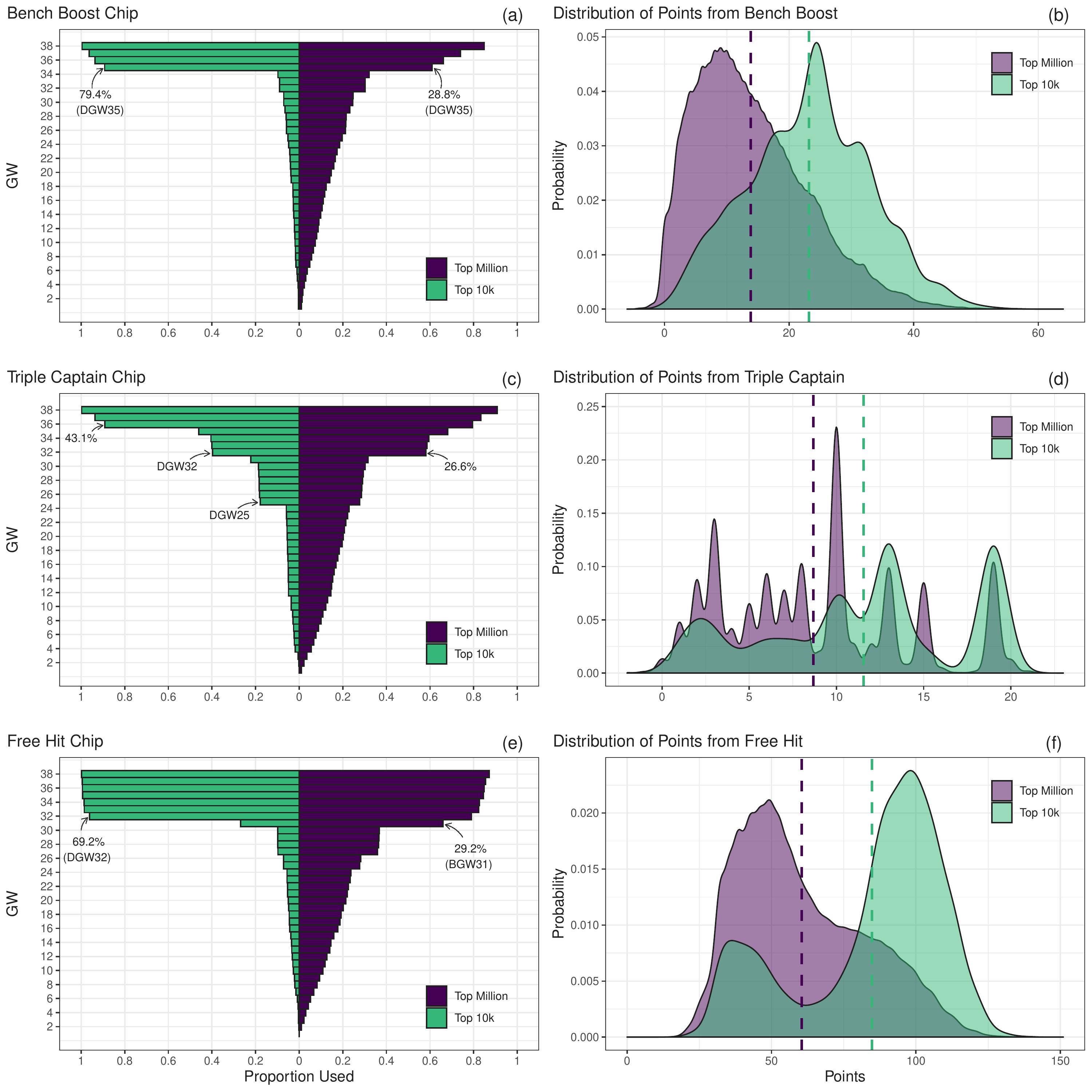}
		\caption{\textbf{Summary results for each of the three chips.} We show the time of the chips' use along with the points received by the manager who played them, for manager who finished in the top $10^3$ and $10^4$ tiers (top 10k) in comparison to the remaining managers (top million) in the dataset. The left panels show the proportion of managers who had used the corresponding chip by each GW (the complementary cumulative distribution function), particularly highlighting the large usages in the `special' gameweeks for each chip. The right panels show the distribution of points received from the chip's use by the two groups of managers, while the mean number of points for each group are also shown by the dashed vertical lines. We comment on the fact that the top 10k received more points on average for each of the three chips.} 
		\label{fig:chip_usage}
	\end{figure*}

	As described in the main text we described three chips which essentially are tricks a manager can make use of in any given gameweek (note that more than one chip can not be used in any single gameweek). The chip properties are summarised below
	\begin{enumerate}
		\item \textbf{Bench Boost (BB)} - The manager receives the points awarded by all 15 players in their squad in comparison to the usual starting 11 players.
		\item \textbf{Free Hit (FH)} - The manager may make unlimited changes to their team for one gameweek, at the end of which their team reverts to the squad from the previous GW, under the standard restriction, i.e., they must remain under their budget, satisfy the formation criterion, and have no more than three players from any one club. 
		\item \textbf{Triple Captain (TC)} - For the GW this chip is played in, the captain's points are tripled rather than doubled. If the captain does not play the triple points are awarded to the vice-captain and, as usual, if neither play no one is awarded triple points.
	\end{enumerate} 
	The points obtained for each of these chips, the distributions of which are shown in \ref{fig:chip_usage}, are calculated by
	\begin{enumerate}
		\item \textbf{Bench Boost (BB)} - We identify the four players on the manager bench the week the chip was played and tally their points total.
		\item \textbf{Free Hit (FH)} - The amount of points the manger received that week is used in this case as the free hit essentially acts like a free week to choose the eleven players of their desire with the aim of maximizing points for one week i.e., no long-term planning is needed.
		\item \textbf{Triple Captain (TC)} - The captain's points total is shown in the distribution. We assume they would have chosen this player to be captain regardless of the chip so would have received double points regardless and as such the difference is only the single points score.
	\end{enumerate} 
	
\begin{figure*}[h!]
		\centering
		\includegraphics[width = 0.75\textwidth]{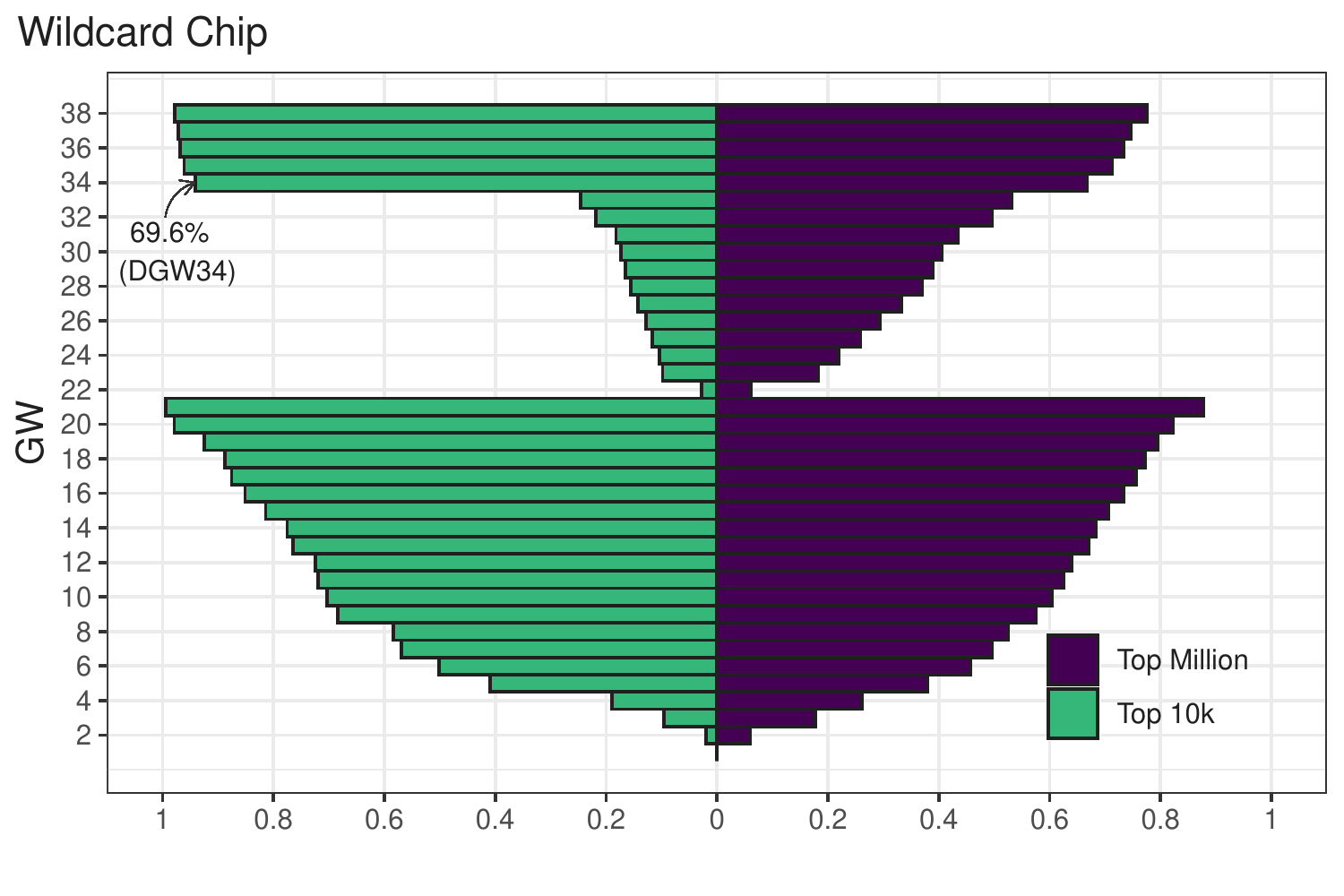}
		\caption{\textbf{Analysis of the wildcard chip's use.} Fraction of managers in the two groups who had used their wildcard chip by each gameweek. Note that the count resets in gameweek 22 when the chip is replenished.} 
		\label{fig:wc_use}
	\end{figure*}

	We repeated this calculation for each gameweek for every manager in our dataset and determined both the number of individuals who played the chip that gameweek alongside the average number of points those that did earned from doing so. The corresponding figures are shown in \ref{tab:BB}, \ref{tab:FH}, and \ref{tab:TC}. A fourth chip also exists in the game and is known as the \textit{wildcard}, this chip allows the manager to make as many transfers as they like in the week it is played thus offering a chance to totally redefine their team. The managers receive this chip twice in the season, the first may only be used between gameweeks 1 and 21, while the second in one of the remaining gameweeks. It proves however much more difficult to quantify the return from this chip e.g., one could consider the wildcarded teams return versus their original team in the following $m$ gameweeks, however in practice the manager would make transfers to their original team in the following gameweeks, another issue is the possibility that the manager `dead-ends' their team up to the week of their wildcard gameweek in the sense that they stop planning for beyond the wildcard opening up the possibility of an extremely biased comparison as the team being changed arguably would not be there without the wildcard. We may still, however, consider the gameweek in which the managers played each of their two chips and this is shown in \ref{fig:wc_use} alongside the quantities themselves in \ref{tab:WC}. Looking at the point of season in which these chips are used we again notice an evident pattern among the actions of the top managers particularly when the second wildcard chip is considered. It appears as though the strategy of choice for those who finished in the top two tiers was to use their free hit during double gameweek 32, wildcard in gameweek 34, and thus having what they believed to be an optimal squad such that they could optimise their bench boost chip which most played in double gameweek 35.
	
	\begin{table*}[ht]
		\centering
		\caption{Usage and Average Points from Bench Boost Chip}
		\label{tab:BB}
		\vspace{-3mm}
		\resizebox{\textwidth}{!}{\begin{tabular}{@{}lcccccccccccc}
				\toprule
				& & {$10^3$} & & & $10^4$ & & & $10^5$ & & & $10^6$ & \\
				\cmidrule(lr){2-4}
				\cmidrule(lr){5-7}
				\cmidrule(lr){8-10}
				\cmidrule(lr){11-13}
				GW & Rel. Freq & Cum. Freq & Mean Points & Rel. Freq & Cum. Freq & Mean Points & Rel. Freq & Cum. Freq & Mean Points & Rel. Ferq & Cum. Freq & Mean Points \\ 
				\midrule
				1 & 0.003 & 0.003 & 20.333 & 0.004 & 0.004 & 21.939 & 0.006 & 0.006 & 16.930 & 0.010 & 0.010 & 11.023 \\ 
				2 & 0.000 & 0.003 & --- & 0.000 & 0.004 & 4.667 & 0.001 & 0.008 & 7.959 & 0.004 & 0.014 & 7.451 \\ 
				3 & 0.000 & 0.003 & --- & 0.000 & 0.004 & 8.000 & 0.001 & 0.009 & 9.907 & 0.004 & 0.018 & 10.022 \\ 
				4 & 0.001 & 0.004 & 12.000 & 0.002 & 0.006 & 8.667 & 0.003 & 0.012 & 9.373 & 0.007 & 0.025 & 8.333 \\ 
				5 & 0.002 & 0.006 & 10.000 & 0.003 & 0.009 & 14.042 & 0.004 & 0.016 & 10.143 & 0.007 & 0.032 & 8.783 \\ 
				6 & 0.000 & 0.006 & --- & 0.001 & 0.010 & 12.500 & 0.003 & 0.019 & 12.431 & 0.007 & 0.039 & 10.543 \\ 
				7 & 0.003 & 0.009 & 13.667 & 0.005 & 0.015 & 12.436 & 0.008 & 0.028 & 11.223 & 0.012 & 0.051 & 10.322 \\ 
				8 & 0.001 & 0.010 & 17.000 & 0.002 & 0.017 & 13.700 & 0.005 & 0.033 & 13.732 & 0.009 & 0.060 & 12.491 \\ 
				9 & 0.000 & 0.010 & --- & 0.001 & 0.018 & 6.375 & 0.004 & 0.037 & 8.076 & 0.008 & 0.068 & 7.717 \\ 
				10 & 0.001 & 0.011 & 2.000 & 0.002 & 0.020 & 6.214 & 0.005 & 0.041 & 7.409 & 0.010 & 0.078 & 7.376 \\ 
				11 & 0.001 & 0.012 & 7.000 & 0.001 & 0.021 & 9.364 & 0.002 & 0.043 & 9.960 & 0.006 & 0.084 & 9.037 \\ 
				12 & 0.000 & 0.012 & --- & 0.001 & 0.023 & 11.083 & 0.003 & 0.046 & 10.810 & 0.007 & 0.091 & 10.317 \\ 
				13 & 0.000 & 0.012 & --- & 0.001 & 0.024 & 12.250 & 0.002 & 0.048 & 9.361 & 0.005 & 0.096 & 8.300 \\ 
				14 & 0.000 & 0.012 & --- & 0.003 & 0.027 & 7.793 & 0.005 & 0.053 & 9.610 & 0.007 & 0.103 & 8.488 \\ 
				15 & 0.001 & 0.013 & 13.000 & 0.001 & 0.028 & 11.222 & 0.003 & 0.056 & 8.466 & 0.007 & 0.110 & 7.018 \\ 
				16 & 0.000 & 0.013 & --- & 0.001 & 0.029 & 6.125 & 0.003 & 0.059 & 8.080 & 0.006 & 0.116 & 6.822 \\ 
				17 & 0.002 & 0.015 & 5.500 & 0.001 & 0.030 & 13.000 & 0.004 & 0.063 & 11.663 & 0.007 & 0.123 & 10.209 \\ 
				18 & 0.000 & 0.015 & --- & 0.001 & 0.031 & 16.444 & 0.002 & 0.065 & 9.754 & 0.006 & 0.129 & 9.492 \\ 
				19 & 0.005 & 0.020 & 16.600 & 0.008 & 0.039 & 15.232 & 0.012 & 0.076 & 11.900 & 0.017 & 0.147 & 10.249 \\ 
				20 & 0.001 & 0.021 & 11.000 & 0.001 & 0.040 & 7.857 & 0.003 & 0.079 & 10.795 & 0.008 & 0.155 & 9.667 \\ 
				21 & 0.002 & 0.023 & 19.000 & 0.002 & 0.042 & 13.333 & 0.006 & 0.085 & 12.410 & 0.011 & 0.166 & 11.187 \\ 
				22 & 0.000 & 0.023 & --- & 0.002 & 0.044 & 11.538 & 0.003 & 0.088 & 10.874 & 0.008 & 0.174 & 10.083 \\ 
				23 & 0.001 & 0.024 & 13.000 & 0.001 & 0.045 & 6.000 & 0.004 & 0.092 & 9.313 & 0.008 & 0.182 & 8.708 \\ 
				24 & 0.004 & 0.028 & 9.250 & 0.005 & 0.051 & 8.200 & 0.010 & 0.102 & 9.220 & 0.012 & 0.194 & 8.208 \\ 
				25 & 0.001 & 0.029 & 12.000 & 0.004 & 0.054 & 14.467 & 0.008 & 0.110 & 14.635 & 0.012 & 0.206 & 13.299 \\ 
				26 & 0.009 & 0.038 & 11.889 & 0.007 & 0.061 & 8.860 & 0.013 & 0.123 & 7.053 & 0.015 & 0.221 & 6.522 \\ 
				27 & 0.001 & 0.039 & 31.000 & 0.000 & 0.061 & 8.000 & 0.000 & 0.123 & 8.387 & 0.001 & 0.223 & 4.789 \\ 
				28 & 0.000 & 0.039 & --- & 0.001 & 0.062 & 12.583 & 0.002 & 0.125 & 12.816 & 0.003 & 0.226 & 10.376 \\ 
				29 & 0.006 & 0.045 & 7.167 & 0.005 & 0.067 & 10.326 & 0.012 & 0.137 & 11.184 & 0.016 & 0.242 & 9.979 \\ 
				30 & 0.004 & 0.049 & 19.250 & 0.006 & 0.073 & 13.922 & 0.010 & 0.147 & 12.906 & 0.014 & 0.256 & 11.599 \\ 
				31 & 0.000 & 0.049 & --- & 0.000 & 0.073 & 8.000 & 0.000 & 0.147 & 5.455 & 0.001 & 0.257 & 3.905 \\ 
				32 & 0.020 & 0.069 & 17.750 & 0.019 & 0.093 & 19.261 & 0.041 & 0.188 & 17.578 & 0.057 & 0.314 & 15.573 \\ 
				33 & 0.000 & 0.069 & --- & 0.000 & 0.093 & 19.000 & 0.000 & 0.189 & 10.320 & 0.001 & 0.315 & 4.628 \\ 
				34 & 0.006 & 0.075 & 16.333 & 0.007 & 0.100 & 15.082 & 0.013 & 0.201 & 12.463 & 0.019 & 0.334 & 10.647 \\ 
				35 & 0.856 & 0.931 & 27.474 & 0.787 & 0.888 & 25.414 & 0.594 & 0.795 & 23.147 & 0.257 & 0.590 & 18.849 \\ 
				36 & 0.033 & 0.964 & 15.000 & 0.045 & 0.933 & 15.592 & 0.052 & 0.847 & 15.775 & 0.052 & 0.642 & 14.832 \\ 
				37 & 0.018 & 0.982 & 13.778 & 0.028 & 0.960 & 13.233 & 0.059 & 0.905 & 12.486 & 0.081 & 0.723 & 11.447 \\ 
				38 & 0.017 & 0.999 & 11.588 & 0.035 & 0.995 & 10.414 & 0.074 & 0.980 & 9.413 & 0.113 & 0.835 & 8.668 \\ 
				\bottomrule
		\end{tabular}}
	\end{table*}
	
	\begin{table*}[ht]
		\centering
		\caption{Usage and Average Points from Free Hit Chip}
		\label{tab:FH}
		\vspace{-3mm}
		\resizebox{\textwidth}{!}{\begin{tabular}{@{}lcccccccccccc}
				\toprule
				& & {$10^3$} & & & $10^4$ & & & $10^5$ & & & $10^6$ & \\
				\cmidrule(lr){2-4}
				\cmidrule(lr){5-7}
				\cmidrule(lr){8-10}
				\cmidrule(lr){11-13}
				GW & Rel. Freq & Cum. Freq & Mean Points & Rel. Freq & Cum. Freq & Mean Points & Rel. Freq & Cum. Freq & Mean Points & Rel. Freq & Cum. Freq & Mean Points \\
				\midrule 
				1 & 0.000 & 0.000 & --- & 0.000 & 0.000 & --- & 0.000 & 0.000 & --- & 0.000 & 0.000 & --- \\ 
				2 & 0.001 & 0.001 & 67.000 & 0.001 & 0.001 & 84.500 & 0.003 & 0.003 & 78.689 & 0.011 & 0.011 & 71.896 \\ 
				3 & 0.000 & 0.001 & --- & 0.002 & 0.003 & 53.308 & 0.004 & 0.007 & 55.192 & 0.012 & 0.023 & 53.375 \\ 
				4 & 0.003 & 0.004 & 56.667 & 0.001 & 0.004 & 62.000 & 0.004 & 0.011 & 53.742 & 0.010 & 0.033 & 51.799 \\ 
				5 & 0.002 & 0.006 & 91.000 & 0.002 & 0.006 & 76.706 & 0.005 & 0.017 & 72.022 & 0.012 & 0.045 & 62.225 \\ 
				6 & 0.001 & 0.007 & 64.000 & 0.002 & 0.008 & 64.250 & 0.004 & 0.021 & 60.878 & 0.010 & 0.056 & 56.928 \\ 
				7 & 0.006 & 0.013 & 79.000 & 0.007 & 0.015 & 81.172 & 0.011 & 0.031 & 76.225 & 0.015 & 0.070 & 68.090 \\ 
				8 & 0.003 & 0.016 & 57.667 & 0.005 & 0.020 & 57.667 & 0.013 & 0.044 & 56.857 & 0.017 & 0.087 & 56.142 \\ 
				9 & 0.000 & 0.016 & --- & 0.002 & 0.022 & 49.214 & 0.005 & 0.049 & 48.288 & 0.011 & 0.098 & 45.351 \\ 
				10 & 0.006 & 0.022 & 97.000 & 0.005 & 0.027 & 88.000 & 0.010 & 0.059 & 83.850 & 0.015 & 0.114 & 78.136 \\ 
				11 & 0.002 & 0.024 & 81.500 & 0.002 & 0.030 & 80.143 & 0.006 & 0.066 & 75.512 & 0.010 & 0.124 & 66.413 \\ 
				12 & 0.004 & 0.028 & 59.250 & 0.002 & 0.032 & 59.062 & 0.004 & 0.070 & 56.708 & 0.009 & 0.133 & 54.179 \\ 
				13 & 0.001 & 0.029 & 90.000 & 0.003 & 0.035 & 69.731 & 0.008 & 0.078 & 66.031 & 0.014 & 0.147 & 61.902 \\ 
				14 & 0.000 & 0.029 & --- & 0.000 & 0.035 & 61.000 & 0.002 & 0.080 & 58.169 & 0.006 & 0.153 & 57.075 \\ 
				15 & 0.005 & 0.034 & 64.400 & 0.004 & 0.039 & 69.581 & 0.007 & 0.087 & 64.119 & 0.013 & 0.166 & 58.872 \\ 
				16 & 0.003 & 0.037 & 64.000 & 0.006 & 0.045 & 78.957 & 0.011 & 0.098 & 72.813 & 0.020 & 0.186 & 65.906 \\ 
				17 & 0.000 & 0.037 & --- & 0.001 & 0.045 & 55.714 & 0.004 & 0.102 & 54.592 & 0.010 & 0.196 & 50.406 \\ 
				18 & 0.000 & 0.037 & --- & 0.002 & 0.047 & 53.231 & 0.003 & 0.105 & 53.367 & 0.006 & 0.202 & 55.985 \\ 
				19 & 0.001 & 0.038 & 94.000 & 0.002 & 0.049 & 84.737 & 0.005 & 0.110 & 80.305 & 0.010 & 0.212 & 74.250 \\ 
				20 & 0.001 & 0.039 & 40.000 & 0.003 & 0.052 & 60.409 & 0.006 & 0.117 & 61.189 & 0.013 & 0.225 & 57.583 \\ 
				21 & 0.003 & 0.042 & 66.333 & 0.002 & 0.054 & 65.429 & 0.004 & 0.121 & 65.541 & 0.006 & 0.231 & 62.016 \\ 
				22 & 0.000 & 0.042 & --- & 0.001 & 0.055 & 62.750 & 0.003 & 0.124 & 62.424 & 0.005 & 0.236 & 58.843 \\ 
				23 & 0.002 & 0.044 & 94.000 & 0.002 & 0.057 & 75.706 & 0.005 & 0.129 & 76.457 & 0.009 & 0.245 & 69.248 \\ 
				24 & 0.000 & 0.044 & --- & 0.000 & 0.057 & 47.667 & 0.001 & 0.130 & 52.558 & 0.003 & 0.248 & 47.562 \\ 
				25 & 0.009 & 0.053 & 94.556 & 0.016 & 0.074 & 91.221 & 0.032 & 0.162 & 88.109 & 0.042 & 0.290 & 82.062 \\ 
				26 & 0.000 & 0.053 & --- & 0.000 & 0.074 & 79.500 & 0.002 & 0.163 & 70.571 & 0.005 & 0.295 & 68.977 \\ 
				27 & 0.016 & 0.069 & 45.000 & 0.027 & 0.101 & 46.240 & 0.058 & 0.221 & 44.662 & 0.079 & 0.373 & 42.860 \\ 
				28 & 0.000 & 0.069 & --- & 0.000 & 0.101 & 79.500 & 0.002 & 0.223 & 69.053 & 0.005 & 0.378 & 65.063 \\ 
				29 & 0.000 & 0.069 & --- & 0.000 & 0.101 & --- & 0.001 & 0.223 & 51.585 & 0.002 & 0.380 & 47.069 \\ 
				30 & 0.000 & 0.069 & --- & 0.000 & 0.101 & 59.000 & 0.001 & 0.224 & 65.154 & 0.002 & 0.382 & 57.933 \\ 
				31 & 0.114 & 0.183 & 39.561 & 0.178 & 0.279 & 40.827 & 0.300 & 0.524 & 42.048 & 0.291 & 0.674 & 43.879 \\ 
				32 & 0.788 & 0.971 & 99.999 & 0.681 & 0.960 & 98.233 & 0.399 & 0.923 & 96.261 & 0.103 & 0.777 & 92.510 \\ 
				33 & 0.021 & 0.992 & 79.762 & 0.025 & 0.985 & 76.359 & 0.037 & 0.961 & 74.060 & 0.033 & 0.809 & 68.152 \\ 
				34 & 0.001 & 0.993 & 36.000 & 0.001 & 0.986 & 56.400 & 0.002 & 0.962 & 54.674 & 0.003 & 0.812 & 54.866 \\ 
				35 & 0.003 & 0.996 & 84.000 & 0.007 & 0.993 & 77.541 & 0.013 & 0.975 & 76.154 & 0.019 & 0.831 & 74.456 \\ 
				36 & 0.001 & 0.997 & 111.000 & 0.001 & 0.993 & 100.000 & 0.002 & 0.977 & 93.881 & 0.005 & 0.835 & 86.754 \\ 
				37 & 0.000 & 0.997 & --- & 0.001 & 0.994 & 63.000 & 0.003 & 0.980 & 56.013 & 0.006 & 0.841 & 55.201 \\ 
				38 & 0.003 & 1.000 & 62.000 & 0.004 & 0.998 & 72.833 & 0.008 & 0.988 & 70.057 & 0.019 & 0.860 & 66.632 \\ 
				\bottomrule
		\end{tabular}}
	\end{table*}	
	
	\begin{table*}[ht]
		\centering
		\caption{Usage and Average Points from Triple Captain Chip}
		\label{tab:TC}
		\vspace{-3mm}
		\resizebox{\textwidth}{!}{\begin{tabular}{@{}lcccccccccccc}
				\toprule
				& & {$10^3$} & & & $10^4$ & & & $10^5$ & & & $10^6$ & \\
				\cmidrule(lr){2-4}
				\cmidrule(lr){5-7}
				\cmidrule(lr){8-10}
				\cmidrule(lr){11-13}
				GW & Rel. Freq & Cum. Freq & Mean Points & Rel. Freq & Cum. Freq & Mean Points & Rel. Freq & Cum. Freq & Mean Points & Rel. Freq & Cum. Freq & Mean Points \\ 
				\midrule
				1 & 0.000 & 0.000 & --- & 0.001 & 0.001 & 8.571 & 0.003 & 0.003 & 7.100 & 0.011 & 0.011 & 5.896 \\ 
				2 & 0.003 & 0.003 & 20.000 & 0.002 & 0.002 & 18.214 & 0.004 & 0.007 & 17.104 & 0.012 & 0.023 & 12.846 \\ 
				3 & 0.003 & 0.006 & 8.000 & 0.003 & 0.006 & 7.000 & 0.006 & 0.013 & 7.015 & 0.014 & 0.037 & 5.971 \\ 
				4 & 0.006 & 0.012 & 6.000 & 0.013 & 0.019 & 6.083 & 0.017 & 0.030 & 6.090 & 0.021 & 0.058 & 5.934 \\ 
				5 & 0.005 & 0.017 & 9.600 & 0.004 & 0.023 & 11.105 & 0.007 & 0.037 & 11.845 & 0.013 & 0.071 & 10.032 \\ 
				6 & 0.001 & 0.018 & 8.000 & 0.001 & 0.024 & 5.000 & 0.003 & 0.039 & 6.167 & 0.010 & 0.081 & 5.868 \\ 
				7 & 0.001 & 0.019 & 8.000 & 0.003 & 0.026 & 8.000 & 0.004 & 0.044 & 8.491 & 0.012 & 0.093 & 8.485 \\ 
				8 & 0.001 & 0.020 & 1.000 & 0.003 & 0.030 & 1.500 & 0.006 & 0.049 & 2.554 & 0.013 & 0.106 & 3.909 \\ 
				9 & 0.000 & 0.020 & --- & 0.001 & 0.031 & 6.222 & 0.002 & 0.052 & 5.387 & 0.008 & 0.114 & 5.094 \\ 
				10 & 0.007 & 0.027 & 15.000 & 0.007 & 0.037 & 15.000 & 0.010 & 0.062 & 14.529 & 0.014 & 0.128 & 13.513 \\ 
				11 & 0.001 & 0.028 & 13.000 & 0.000 & 0.038 & 15.000 & 0.003 & 0.065 & 12.974 & 0.009 & 0.137 & 11.468 \\ 
				12 & 0.010 & 0.038 & 8.000 & 0.012 & 0.050 & 7.670 & 0.013 & 0.078 & 7.538 & 0.015 & 0.152 & 6.611 \\ 
				13 & 0.001 & 0.039 & 3.000 & 0.000 & 0.050 & 3.000 & 0.001 & 0.079 & 6.391 & 0.004 & 0.156 & 5.252 \\ 
				14 & 0.000 & 0.039 & --- & 0.000 & 0.051 & 6.500 & 0.002 & 0.081 & 4.089 & 0.005 & 0.161 & 3.565 \\ 
				15 & 0.000 & 0.039 & --- & 0.001 & 0.051 & 12.000 & 0.003 & 0.083 & 9.582 & 0.007 & 0.169 & 7.424 \\ 
				16 & 0.000 & 0.039 & --- & 0.001 & 0.052 & 4.500 & 0.003 & 0.086 & 5.891 & 0.010 & 0.178 & 6.045 \\ 
				17 & 0.001 & 0.040 & 5.000 & 0.001 & 0.054 & 4.300 & 0.003 & 0.089 & 6.175 & 0.008 & 0.186 & 5.549 \\ 
				18 & 0.000 & 0.040 & --- & 0.001 & 0.055 & 4.625 & 0.002 & 0.091 & 6.882 & 0.008 & 0.194 & 7.408 \\ 
				19 & 0.001 & 0.041 & 6.000 & 0.001 & 0.056 & 11.500 & 0.004 & 0.096 & 10.410 & 0.014 & 0.208 & 10.073 \\ 
				20 & 0.000 & 0.041 & --- & 0.001 & 0.057 & 10.286 & 0.002 & 0.098 & 9.729 & 0.006 & 0.214 & 8.041 \\ 
				21 & 0.000 & 0.041 & --- & 0.001 & 0.058 & 8.250 & 0.002 & 0.100 & 8.131 & 0.005 & 0.219 & 7.467 \\ 
				22 & 0.000 & 0.041 &--- & 0.000 & 0.058 & 8.000 & 0.001 & 0.101 & 8.686 & 0.006 & 0.225 & 8.116 \\ 
				23 & 0.000 & 0.041 & --- & 0.002 & 0.060 & 13.929 & 0.003 & 0.104 & 12.465 & 0.009 & 0.234 & 12.138 \\ 
				24 & 0.001 & 0.042 & 2.000 & 0.001 & 0.061 & 7.182 & 0.003 & 0.107 & 6.650 & 0.007 & 0.241 & 6.418 \\ 
				25 & 0.152 & 0.194 & 14.066 & 0.115 & 0.176 & 12.720 & 0.077 & 0.184 & 12.068 & 0.046 & 0.287 & 12.989 \\ 
				26 & 0.005 & 0.199 & 8.000 & 0.004 & 0.180 & 8.000 & 0.006 & 0.190 & 7.874 & 0.009 & 0.296 & 7.662 \\ 
				27 & 0.001 & 0.200 & 2.000 & 0.001 & 0.181 & 1.400 & 0.000 & 0.191 & 2.310 & 0.001 & 0.297 & 2.208 \\ 
				28 & 0.000 & 0.200 & --- & 0.001 & 0.182 & 5.571 & 0.002 & 0.193 & 5.558 & 0.004 & 0.301 & 5.509 \\ 
				29 & 0.001 & 0.201 & 3.000 & 0.001 & 0.183 & 2.500 & 0.002 & 0.195 & 2.947 & 0.004 & 0.305 & 2.719 \\ 
				30 & 0.002 & 0.203 & 5.000 & 0.002 & 0.185 & 10.056 & 0.004 & 0.199 & 7.042 & 0.009 & 0.314 & 6.591 \\ 
				31 & 0.027 & 0.230 & 2.000 & 0.036 & 0.221 & 2.581 & 0.025 & 0.224 & 3.083 & 0.011 & 0.324 & 4.741 \\ 
				32 & 0.121 & 0.351 & 9.405 & 0.182 & 0.403 & 9.370 & 0.291 & 0.516 & 9.093 & 0.264 & 0.588 & 8.451 \\ 
				33 & 0.002 & 0.353 & 16.000 & 0.004 & 0.406 & 14.387 & 0.004 & 0.520 & 12.335 & 0.005 & 0.593 & 8.983 \\ 
				34 & 0.001 & 0.354 & 1.000 & 0.004 & 0.411 & 2.216 & 0.007 & 0.526 & 2.938 & 0.008 & 0.601 & 3.580 \\ 
				35 & 0.056 & 0.410 & 6.893 & 0.056 & 0.467 & 6.686 & 0.077 & 0.603 & 6.570 & 0.088 & 0.689 & 6.469 \\ 
				36 & 0.505 & 0.915 & 15.986 & 0.422 & 0.889 & 15.376 & 0.266 & 0.869 & 15.136 & 0.098 & 0.787 & 14.586 \\ 
				37 & 0.038 & 0.953 & 4.368 & 0.046 & 0.935 & 4.205 & 0.045 & 0.914 & 4.106 & 0.039 & 0.826 & 4.165 \\ 
				38 & 0.045 & 0.998 & 6.111 & 0.062 & 0.997 & 6.779 & 0.074 & 0.988 & 6.781 & 0.075 & 0.900 & 6.796 \\ 
				\bottomrule
		\end{tabular}}
	\end{table*}	
	
	\begin{table*}[ht]
		\centering
		\caption{Usage of Wildcard Chip}
		\label{tab:WC}
		\vspace{-3mm}
		\resizebox{\textwidth}{!}{\begin{tabular}{@{}lcccccccc}
				\toprule
				& \multicolumn{2}{c}{$10^3$} & \multicolumn{2}{c}{$10^4$} & \multicolumn{2}{c}{$10^5$} & \multicolumn{2}{c}{$10^6$} \\
				\cmidrule(lr){2-3}
				\cmidrule(lr){4-5}
				\cmidrule(lr){6-7}
				\cmidrule(lr){8-9}
				GW & Rel. Freq & Cum. Freq & Rel. Freq & Cum. Freq & Rel. Freq & Cum. Freq & Rel. Freq & Cum. Freq \\ 
				\midrule
				1 & --- & --- & --- & --- & --- & --- & --- & --- \\ 
				2 & 0.019 & 0.019 & 0.020 & 0.020 & 0.040 & 0.040 & 0.062 & 0.062 \\ 
				3 & 0.055 & 0.074 & 0.078 & 0.098 & 0.104 & 0.143 & 0.119 & 0.181 \\ 
				4 & 0.093 & 0.167 & 0.094 & 0.192 & 0.092 & 0.235 & 0.083 & 0.264 \\ 
				5 & 0.233 & 0.400 & 0.218 & 0.410 & 0.187 & 0.422 & 0.111 & 0.375 \\ 
				6 & 0.071 & 0.471 & 0.095 & 0.505 & 0.091 & 0.513 & 0.076 & 0.451 \\ 
				7 & 0.083 & 0.554 & 0.067 & 0.571 & 0.054 & 0.567 & 0.037 & 0.489 \\ 
				8 & 0.011 & 0.565 & 0.014 & 0.585 & 0.025 & 0.593 & 0.029 & 0.518 \\ 
				9 & 0.103 & 0.668 & 0.100 & 0.686 & 0.075 & 0.668 & 0.048 & 0.566 \\ 
				10 & 0.014 & 0.682 & 0.020 & 0.705 & 0.029 & 0.697 & 0.028 & 0.594 \\ 
				11 & 0.016 & 0.698 & 0.016 & 0.721 & 0.017 & 0.714 & 0.022 & 0.616 \\ 
				12 & 0.004 & 0.702 & 0.006 & 0.728 & 0.010 & 0.724 & 0.014 & 0.630 \\ 
				13 & 0.031 & 0.733 & 0.040 & 0.768 & 0.038 & 0.763 & 0.030 & 0.661 \\ 
				14 & 0.012 & 0.745 & 0.011 & 0.779 & 0.011 & 0.774 & 0.013 & 0.674 \\ 
				15 & 0.052 & 0.797 & 0.037 & 0.816 & 0.032 & 0.806 & 0.021 & 0.695 \\ 
				16 & 0.045 & 0.842 & 0.037 & 0.852 & 0.037 & 0.843 & 0.027 & 0.722 \\ 
				17 & 0.029 & 0.871 & 0.024 & 0.876 & 0.022 & 0.865 & 0.023 & 0.745 \\ 
				18 & 0.009 & 0.880 & 0.012 & 0.888 & 0.014 & 0.879 & 0.016 & 0.760 \\ 
				19 & 0.037 & 0.917 & 0.037 & 0.925 & 0.028 & 0.907 & 0.023 & 0.783 \\ 
				20 & 0.075 & 0.992 & 0.052 & 0.977 & 0.038 & 0.945 & 0.026 & 0.809 \\ 
				21 & 0.009 & 1.001 & 0.016 & 0.993 & 0.034 & 0.980 & 0.057 & 0.866 \\ 
				22 & 0.016 & 0.016 & 0.029 & 0.029 & 0.052 & 0.052 & 0.063 & 0.063 \\ 
				23 & 0.050 & 0.066 & 0.072 & 0.102 & 0.112 & 0.164 & 0.122 & 0.185 \\ 
				24 & 0.002 & 0.068 & 0.007 & 0.109 & 0.019 & 0.182 & 0.039 & 0.223 \\ 
				25 & 0.005 & 0.073 & 0.014 & 0.122 & 0.028 & 0.210 & 0.040 & 0.264 \\ 
				26 & 0.008 & 0.081 & 0.011 & 0.134 & 0.022 & 0.233 & 0.036 & 0.300 \\ 
				27 & 0.014 & 0.095 & 0.015 & 0.148 & 0.026 & 0.259 & 0.041 & 0.341 \\ 
				28 & 0.008 & 0.103 & 0.014 & 0.162 & 0.029 & 0.288 & 0.037 & 0.378 \\ 
				29 & 0.008 & 0.111 & 0.009 & 0.171 & 0.016 & 0.303 & 0.020 & 0.398 \\ 
				30 & 0.005 & 0.116 & 0.009 & 0.180 & 0.013 & 0.316 & 0.017 & 0.415 \\ 
				31 & 0.003 & 0.119 & 0.009 & 0.189 & 0.019 & 0.336 & 0.030 & 0.445 \\ 
				32 & 0.032 & 0.151 & 0.037 & 0.227 & 0.064 & 0.399 & 0.061 & 0.506 \\ 
				33 & 0.026 & 0.177 & 0.027 & 0.254 & 0.041 & 0.440 & 0.035 & 0.541 \\ 
				34 & 0.793 & 0.970 & 0.684 & 0.938 & 0.420 & 0.861 & 0.106 & 0.647 \\ 
				35 & 0.011 & 0.981 & 0.021 & 0.959 & 0.040 & 0.900 & 0.046 & 0.693 \\ 
				36 & 0.004 & 0.985 & 0.007 & 0.966 & 0.014 & 0.914 & 0.021 & 0.714 \\ 
				37 & 0.002 & 0.987 & 0.004 & 0.970 & 0.008 & 0.922 & 0.014 & 0.728 \\ 
				38 & 0.002 & 0.989 & 0.008 & 0.977 & 0.017 & 0.939 & 0.030 & 0.758 \\ 
				\bottomrule
		\end{tabular}}
	\end{table*}

\end{document}